\documentclass[acmsmall,screen]{acmart}

\usepackage{titlesec}
\usepackage{array}
\usepackage{hyperref}
\usepackage{multirow}
\usepackage{array} 
\usepackage{enumitem}
\usepackage{tikz} 
\usepackage{subcaption}
\usepackage{amsmath}
\usepackage{longtable}
\usepackage{lscape}
\usepackage{graphicx}
\tikzset{font={\fontsize{8pt}{10}\selectfont}}

\usepackage{dsfont} 

\AtBeginDocument{%
  }

\acmDOI{
}


\begin{document}

\title{Privacy risk in GeoData: A survey}

\author{Mahrokh Abdollahi Lorestani}
\email{mahrokh.abdollahilorestani@data61.csiro.au}
\orcid{1234-5678-9012}
\affiliation{%
    \institution{Data61, CSIRO}
    \country{Australia}
}

\author{Thilina Ranbaduge}
\email{thilina.ranbaduge@data61.csiro.au}
\orcid{1234-5678-9012}
\affiliation{%
    \institution{Data61, CSIRO}
    \country{Australia}
  }

\author{Thierry Rakotoarivelo}
\email{thierry.rakotoarivelo@data61.csiro.au}
\affiliation{%
    \institution{Data61, CSIRO}
    \country{Australia}
}
\renewcommand{\shortauthors}{Lorestani et al.}

\begin{abstract}
With the ubiquitous use of location-based services, large-scale 
individual-level location data has been widely collected through 
location-awareness devices. The widespread exposure of such location data poses significant privacy risks to users, as it can lead to re-identification, the inference of sensitive information, and even physical threats. In this survey, we analyse different geomasking techniques proposed to protect individuals' privacy in geodata. We propose a taxonomy to characterise these techniques across various dimensions. We then highlight the shortcomings of current techniques and discuss avenues for future research. Our proposed taxonomy serves as a practical resource for data custodians, offering them a means to navigate the extensive array of existing privacy mechanisms and to identify those that align most effectively with their specific requirements. 
\end{abstract}


\maketitle

\section{Introduction}
\label{sec:introduction}

Data privacy refers to individuals' protection and control over their personal data, 
particularly concerning its collection, processing, storage, and sharing by 
organisations \cite{houser2018gdpr,rustad2019towards}. It is crucial for any data 
that pertains to, is caused by, or initiated by people. 
Organisations across various business sectors increasingly produce large databases
with millions of records, potentially containing detailed and sensitive 
information about individuals, such as customers, patients, taxpayers, or travellers~\cite{drechsler2021synthesizing}. Consequently, 
safeguarding the privacy of such data has become a paramount concern.

%

Privacy violation occurs when personal information is utilised beyond permissible
norms. Thus, the concept of data privacy serves as a guiding principle dictating
the collection and handling of data based on its sensitivity and significance. 
However, while contextual integrity elucidates how privacy breaches can manifest
within a dataset by analysing adherence to social norms specific to different 
contexts, it lacks protective mechanisms beyond policies and 
regulation~\cite{nissenbaum2004}.

Geospatial data (referred to herein as geodata)  is recognised as crucial 
information across various domains, including epidemiology, medicine, and
social science~\cite{fronterre2018}. At its most detailed level, this data
comprises coordinates pinpointing specific locations. For example, geodata
can be linked with other information to create geocoded data, which maps 
individuals' address details into geographic coordinates. The amount of 
high-resolution spatial data has been growing enormously due to the 
integration of positioning capabilities into mobile, wearable, and global
positioning systems. With the increasing availability of data about 
individuals, geodata enables the precise identification of individuals’ 
locations~\cite{sharad2013, de2013, unnikrishnan2013}. 

As geodata become increasingly accessible at finer resolutions, 
attacker capabilities in linking geodata to specific individuals or 
groups continue to grow. Knowing an individual’s location increases the 
risk of reidentification through reverse geocoding and can seriously violate
the individual’s privacy~\cite{brownstein2006, kounadi2013}. For instance,
when publishing a data point, coordinates can be simply calculated and 
connected with an address. Therefore, the address attribute should be 
treated as confidential information because it can reveal an individual.

The growing need for protecting discrete geographic data results from 
technological developments and the need to release data online. Hence, 
numerous studies in the past decade have proposed various approaches to 
preserve sensitive information, such as \emph{geomasking} techniques~\cite{zandbergen2014ensuring}. Geomasking techniques are meant
to preserve privacy when publishing geodata while maintaining geographic
detail to enable precise spatial analysis of the dataset~\cite{kwan2004protection,gambs2014,wang2022exploratory}. These 
techniques aim to obfuscate the real locations of an individual. For 
instance, by employing geomasking techniques, the address of an individual
does not have to be eliminated but can be transferred to a different 
position, which would result in less accurate data utility but higher privacy.

Despite the various techniques proposed, there are no generally 
recommended nor approved geomasking techniques. Each presents its 
unique set of advantages and disadvantages. Furthermore, none of the 
current masking methods offer a comprehensive solution to protect 
location privacy.  Therefore, a thorough analysis of the capabilities
of each technique is imperative to harness them effectively in practical 
applications. Moreover, ongoing research and refinement are necessary
to enhance the efficacy of geomasking strategies in preserving privacy.

\textbf{Comparison to other works:}
Several surveys have studied the geomasking techniques.  
Steffen et al.~\cite{steffenoverview} systematically review the existing
literature on data protection strategies for geocoded data and classify
them into aggregation, geographic masking, and synthetic data. The survey
also discusses various tools for assessing the risk and utility of 
protected geocoded data.  The recent thesis by Redlich~\cite{redlichquantitative}
presents a comprehensive evaluation of geomasking methods designed to 
balance individual privacy protection with the preservation of spatial
information utility. This work provides valuable insights into 
geographic masking methods' strengths, limitations, and privacy 
implications, forming a basis for informed decision-making in spatial
data analysis. Wang et al.~\cite{wang2022exploratory} conduct a 
comprehensive evaluation of eight geomasking methods using simulated 
geospatial data with diverse spatial patterns to assess their 
effectiveness in privacy protection and analytical accuracy. 

Some studies have explored the applicability of geomasking techniques
in specific domains. Kamel Boulos et al.~\cite{kamel2022reconciling} provide
a comprehensive overview of the challenges surrounding location privacy
and techniques for preserving geoprivacy within the realm of public
health interventions and health research that utilise detailed individual
geographic data.  
\cite{iyer2023advancing, broen2021measuring,zandbergen2014ensuring}
explore the use of geomasking techniques for perturbing spatial 
statistics and patient privacy in epidemiologic research and health 
analytics. 
Nowbakht et al.~\cite{nowbakht2022comparison} investigate the challenge
of balancing confidentiality and spatial pattern preservation in sharing
agricultural data. The thesis by Kekana~\cite{kekana2020way} focuses on evaluating
geomasking methods on
presenting the sleeping locations of the homeless without compromising 
their spatial confidence. Gao et al.\cite{gao2019exploring} examine the 
performance of random and Gaussian perturbation methods for Twitter users
sharing their geotagged tweets. 

Further, different studies address a specific concern within the geoprivacy 
domain~\cite{solymosi2023}. Seidl et al.~\cite{seidl2018privacy} explore the 
assessment of the 
risk of false identification in geomasking techniques. Dupre et al.~\cite{dupregeospatial} 
propose to manage privacy risks associated with geocoded data through 
geospatial disclosure avoidance techniques while Tiwari et al.~\cite{tiwari2023exploring} 
investigate the ethical utilisation of geographic information system (GIS) data 
using geomaking techniques. Kim et al.~\cite{kim2021people} investigate 
how do different geomasking methods impact perceived disclosure risk 
and Hu~\cite{hu2018bayesian} explore the utilisation of Bayesian methods 
for estimating disclosure risks in synthetic data. 

Overall, all the proposed studies discuss the different geomasking techniques 
and challenges associated with the increasing availability of geocoded data.
To the best of our knowledge, there has not yet been a thorough survey of 
geomasking techniques and classify these techniques under various aspects. 
This article aims to fill that gap. 

\textbf{Contribution:} We present an extensive survey of 
geomasking techniques. We introduce a taxonomy
that reviews geomasking methods across various dimensions, namely privacy
techniques, evaluation measures, and practical considerations. At a high 
level, this taxonomy differentiates between geomasking techniques based on
how each technique can anonymise geodata, the diverse evaluation measures
applicable to them, and practical aspects. These practical aspects encompass
the experimental datasets, purpose-built prototypes designed specifically for 
implementing these techniques, and the wide range of application domains 
where these techniques find relevance. The goal of this taxonomy is to 
facilitate data custodians in navigating the extensive array of existing
mechanisms, aiding them in identifying those best suited to their needs. 
Furthermore, this survey endeavors to highlight any existing gaps in these 
techniques, offering valuable insights for future researchers to explore 
and refine. While our survey provides guidance across various aspects, we 
acknowledge that other taxonomies proposed by different studies could have
been equally adopted. Some of these taxonomies may offer more detailed 
insights for specific inquiries (e.g., scientists interested in exploring 
knowledge gaps in the field).
%
Additionally, we offer an overview of advanced large 
language models (LLMs) specifically designed to tackle geospatial data 
analytics. While these LLMs may not directly address privacy issues, they 
provide a valuable framework for researchers to explore and integrate 
privacy considerations, enabling  them to address complex geoprivacy 
issues within their methodologies.

\textbf{Outline: } The rest of this paper is structured as follows. First, 
in the following section, 
we provide an overview of geomasking where we discuss different types of geodata and
different privacy attacks on geodata. In Section~\ref{sec:taxonomy} we then present 
a taxonomy on geomasking where we describe the 10 dimensions we identified that 
allow us to characterise geomasking techniques. In Section~\ref{sec:survey} we then 
survey existing geomasking techniques, and describe how they fit into our taxonomy. 
We then discuss open issues in existing geomasking techniques and directions for 
future research in Section~\ref{sec:future-directions}, 
and we conclude this paper in Section~\ref{sec:conclusion} with a summary of our
findings.

\section{Background}
\label{sec:background}

This section provides necessary background details on geomasking, starting with
different types of geodata and then an investigation into different attacks on
geodata. 

\subsection{Different types of Geodata}
Geodata comprises information describing objects, events, or features
with a location on Earth, recorded alongside a geographic indicator. 
It primarily exists in two forms: vector data and raster data. Vector
data delineates features such as properties, cities, roads, mountains, 
and bodies of water, represented by points, lines, and polygons. For 
instance, a visual representation using vector data might include houses
represented by points, roads by lines, and entire towns by polygons. 
Conversely, raster data consists of pixelated or gridded cells, creating
complex imagery like photographs and satellite images.

Besides vector and raster data, location information manifests in various 
other forms, including census data, mobility data, drawn images, and 
social media data. Census data provides insights into specific 
geographical areas, facilitating statistical analysis across different 
locations. Mobility data offers insights into individuals' locations and 
movements, which can be collected through tagging locations on social 
media, using location-tracking apps, interacting with WiFi beacons, or
through Bluetooth, Global Positioning System (GPS), or mobile phone 
records. This data is valuable for conducting geospatial analysis.
Drawn images, such as Computer-aided design (CAD) images, 
provide information about buildings or other structures, offering
geographical insights. Furthermore, social media data can include 
information about people's movements as they post about their visited places.

\subsection{Different Privacy attacks on Geodata}

Geodata privacy refers to a set of principles and practices put in place
to protect the privacy of individuals, groups, communities or organisations whose data is represented in geographically explicit formats~\cite{kamel2022reconciling}. With geodata increasingly 
available at finer resolutions and attackers continuously advancing
their capabilities in linking geodata to specific individuals or
groups, the risk of a privacy breach escalates, especially when 
precise geographic locations are known, particularly in areas with 
low population density. The emergence of various geodata sources
introduces new concerns for privacy breaches, as traditional 
disclosure avoidance mechanisms may overlook spatial 
characteristics~\cite{bertino2008,solymosi2023}.
In this section, we explore a range of privacy attacks/breaches 
targeting geodata. 

\begin{figure}[!t]
    \centering
    \includegraphics[width= 1.0\linewidth, keepaspectratio]{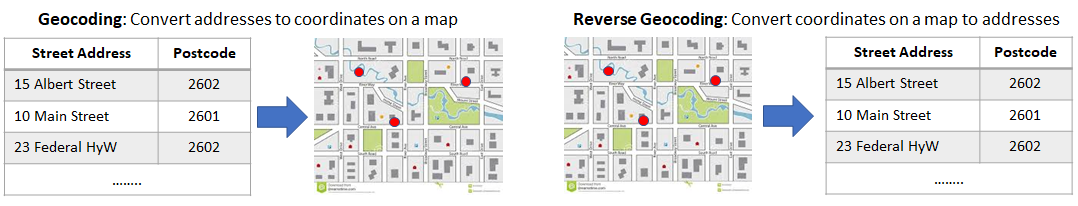}
    \caption{Fundamental process of address geocoding and reverse address geocoding.}
    \label{fig:geocoding}
\end{figure}

\subsubsection{Reverse Geocoding}
~\\
Geocoding is the process of assigning geographic coordinates to addresses or place names, enabling spatial data analysis and visualisation. However, releasing geographic information at the individual level can pose significant privacy risks. For instance, disclosing a street address can allow individuals to be easily identified through directories and property databases. Similarly, publishing coordinates enables pinpoint locations on maps, potentially revealing sensitive information. Whether in print or digital format, maps containing such data can inadvertently disclose addresses and compromise privacy.

Reverse geocoding occurs when the address details for a published location are determined. These address details serve as powerful keys for linking the data associated with those coordinates to specific and identifiable information from other sources. For example, in the context of census data, the respondents' locations can be inferred and reverse geocoded if a census area has cells with low values, potentially leading to inadvertent disclosure.
Reverse geocoding can facilitate the re-identification of individuals, as it allows the conversion of geographic coordinates into street addresses. Once a street address is obtained, it becomes possible to use common directories to associate that address with one or several individuals. Figure~\ref{fig:geocoding} illustrates an example of geocoding and reverse geocoding processes for addresses~\cite{kounadi2013}.

Several research works have shown the re-identification of individual addresses using reverse geocoding. In 2006, Brownstein and others~\cite{brownstein2006} were able to 
correctly identify more than 70\% of the addresses in a published map using manual 
reverse geocoding techniques in GIS. The same authors showed similar accuracy in identifying
addresses can be achieved using semi-automated reverse geocoding based on image 
analysis~\cite{brownstein2006b}. In realistic examples, a group of researchers from the US correctly identified most of the original residents of mortality locations of Hurricane Katrina using a published map in a local newspaper~\cite{curtis2006}.
Further, another study of crime incidents in Vienna, Austria, showed that individuals 
could be accurately re-identified using online reverse geocoding and online address and 
telephone directories~\cite{kounadi2013}.

\subsubsection{Geocoding of administratively-masked data}
~\\
Helderop et al.~\cite{helderop2023unmasking} propose a geocoding 
method for unmasking administratively-masked data where specific identifying 
features, like addresses, are obscured to protect privacy. The proposed approach
utilises a medoid-based technique, aiming to minimise spatial uncertainty associated 
with the masking process. The method utilises a master address point dataset 
to generate a list of candidate addresses for each masked observation, and a
minimum bounding polygon is drawn around these candidates. The medoid, 
representing the central-most point, is then identified within this polygon
and assigned as the geocoded address. The approach outperforms commercial 
geocoding software, providing higher accuracy and a spatial confidence metric
for each result. 

\subsubsection{Inference attack}
~\\
An inference attack corresponds to a process by which an adversary that has access 
to some data related to individuals (and potentially some auxiliary information) 
tries to deduce new personal information that was not explicitly present in the 
original data. For geographical data, the inference attacks have been studied 
for a long time~\cite{sharad2013, de2013, unnikrishnan2013}. The most common inference 
attack type applied on geodata is de-anonymising attacks~\cite{gambs2014}. The de-anonymising 
attack aims to re-identify individuals using their anonymised location data. 

Gambs et al.~\cite{gambs2014} introduce a de-anonymising attack strategy, where an adversary tries to infer the identity of a specific individual through analysis of their mobility traces. More precisely, this attack method demonstrates a high success rate in re-identifying individuals whose movements are documented in an anonymous dataset. This success is contingent upon the adversary having access to mobility traces of the same individuals, observed during a training phase, which they can employ as background information.

Wang et al.~\cite{wang2016} propose a new model to profile users’ spatiotemporal mobility 
behaviours and use the model to launch the de-anonymisation attack. The authors use the User Hidden Markov Model to profile the user movements and used an algorithm to rank the users 
based on the probability that each individual matches with a given profile.

A recent work by Eshun and Palmieri~\cite{eshun2022} proposes two de-anonymisation techniques using 
Hidden Markov Models. In their attacks, both the temporal and spatial influences on user 
mobility trajectories are considered for predicting user movements. These
predictions are then used to create a profile for each individual, and the re-identification of individuals based on these user profiles is achieved through a divergence measure.
An experimental evaluation on real-world mobility traces shows that their proposed 
attack can achieve over 80\% of accuracy in re-identifying individuals.

\subsubsection{Differencing attack}
~\\
A differencing attack is possible with variables for which there are multiple different 
plausible coding schemes, where the categories in those coding schemes are not nested
but instead overlap. For example, given a table with information on 20- to 25-year-olds and a table with information on 20- to 24-year-olds, the difference between the two tables will reveal information about 25-year-olds only. If the difference between cells contains a small number of individuals, then a disclosure is more likely. 

Differencing attacks occur when data is released according to a geospatial hierarchy
such that data in different granularity levels can be combined to reconstruct data 
at a finer granularity or identify the location of an observation. This situation 
may occur for maps with different geographical coding potentially allowing more 
information to be revealed about individuals in the overlaps than intended from 
each individual table. Although it could happen with tabular data, the issue most 
commonly comes up with geodata due to the publicly available geospatial 
information. The result of this differencing is that whilst a 
single map may be considered safe in isolation, this may not be the case for
multiple maps when overlain with one another.

\section{A Taxonomy of Geomasking}
\label{sec:taxonomy}
\begin{figure*}[!t]
    \centering
    \includegraphics[width= 0.8\textwidth, keepaspectratio]{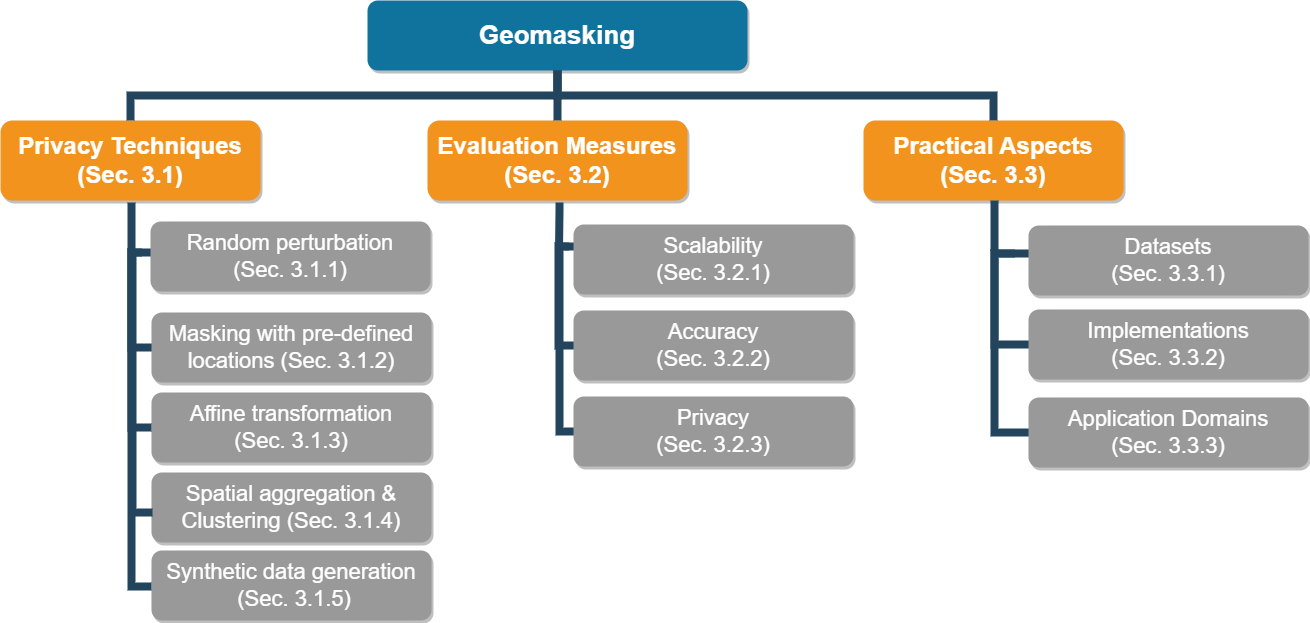}
    \caption{Taxonomy of geomasking.}
    \label{fig:taxonomy}
\end{figure*}

In this section, we outline our proposed taxonomy 
encompassing three primary dimensions of geomasking techniques: privacy 
techniques, evaluation measures, and practical aspects, as illustrated 
in Fig.~\ref{fig:taxonomy}. As previously discussed in 
Section \ref{sec:introduction}, 
our aim in developing this taxonomy is to thoroughly analyse
existing geomasking techniques, the measures to evaluate
their performance, potential datasets for experimental evaluations, 
existing purpose-built prototypes of the techniques, the diverse domains
where these techniques can be applied, and to identify gaps in existing
geomasking techniques. 
Such analysis will assist data custodians in navigating existing methods 
and prototypes to identify the most suitable ones based on their 
specific requirements. In the following subsections, we discuss each 
dimension in detail.

\subsection{Privacy techniques for Geomasking}
\label{sec:geomasking-techniques}

Geomasking is a class of methods for changing the geographic location of an 
individual in an unpredictable way to protect confidentiality, while trying
to preserve the relationship between geocoded locations and occurrence events. 
%
In this section, we analyse different privacy techniques proposed for geomasking. 
Following earlier studies~\cite{armstrong1999geographically,stinchcomb2004procedures,
zandbergen2014ensuring,seidl2018privacy,kekana2020way,broen2021measuring,
wang2022exploratory,redlichquantitative,kamel2022reconciling,
iyer2023advancing}, we group geomasking techniques into five major categories
based on how each technique is treating the data points to provide privacy:
(1) \emph{Random perturbation}, (2) \emph{Masking with pre-defined locations},
(3) \emph{Affine transformation}, (4) \emph{Aggregation and clustering}, and (5) \emph{Synthetic data generation}. Next, we explore different
techniques that are proposed under each of these categories and discuss their
potential and limitations. The existing literature for each category is further 
discussed in Section \ref{sec:survey}.

\subsubsection{Random Perturbation}
~\\
\noindent Random perturbation moves individual points 
from their original location to a random distance in a random 
direction~\cite{allshouse2010geomasking}. When the generated masked points 
may be close to the original points and for a small proportion of the geomasked
points, the adversary can re-identify individuals by reversing geocode. 
Next, we explain several techniques that can be categorised under random 
perturbation. 

\noindent\paragraph{Density-adaptive displacement}
~\\
In this category, the displacement distance is inversely proportional to the 
underlying population density. In the geomasking literature, the population
densities can be considered either as \emph{uniform} or \emph{heterogeneous}.

\begin{figure*}[!t]
    \centering
    \includegraphics[width= 0.3\linewidth, keepaspectratio]{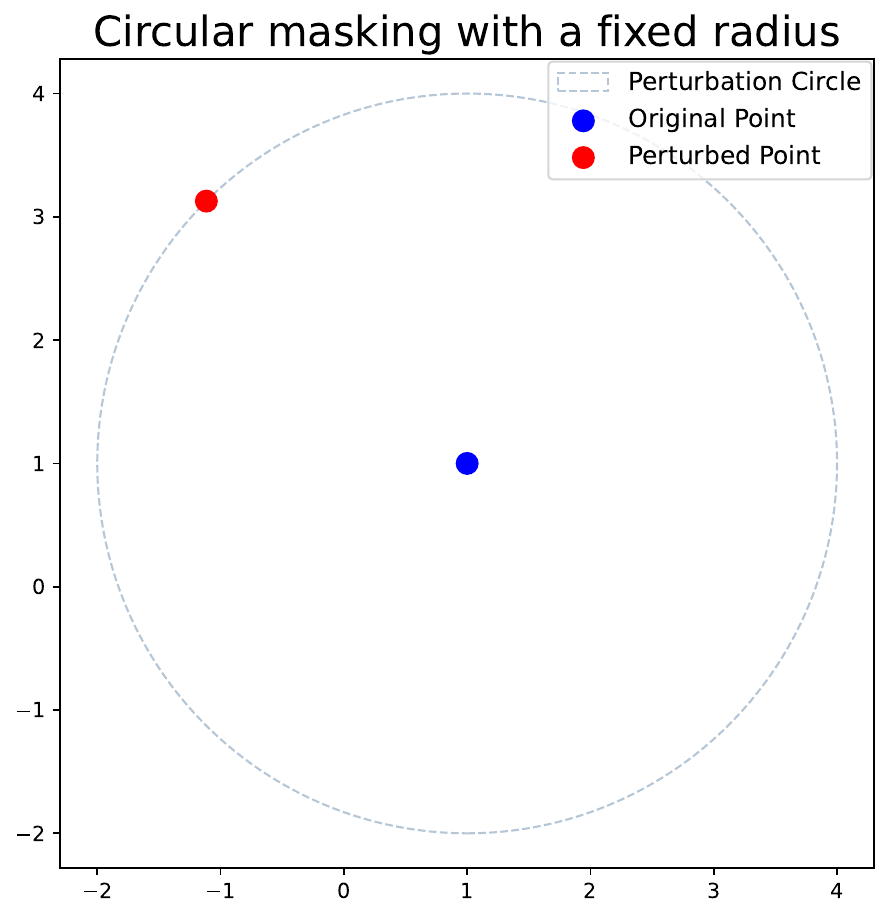}
    \includegraphics[width= 0.3\linewidth, keepaspectratio]{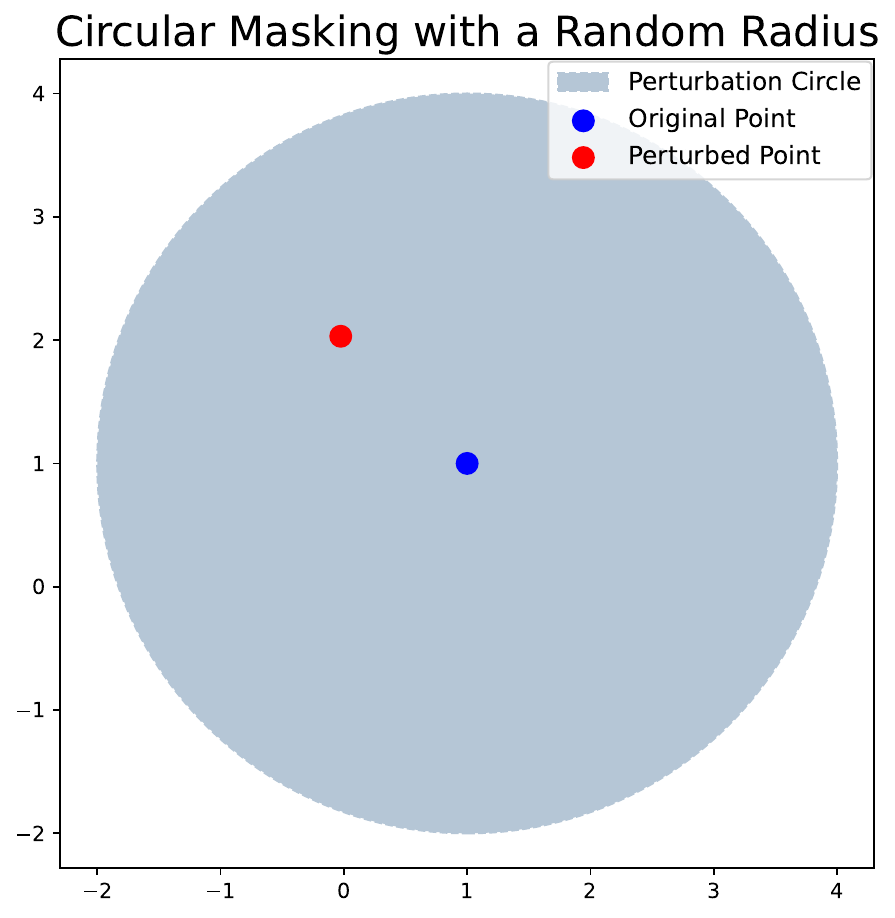}
    \includegraphics[width= 0.3\linewidth, keepaspectratio]{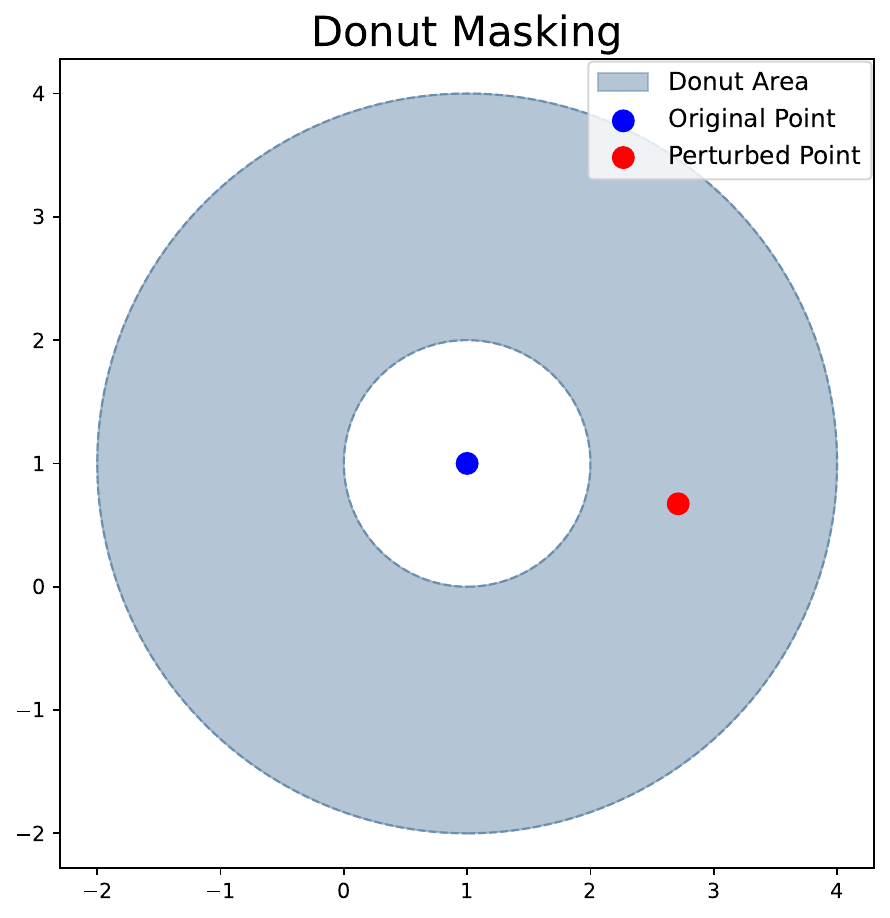}
    \caption{Circular masking with fixed radius (left), 
    Circular masking with random radius (middle),
    and donut masking (right).}
    \label{fig:cir_mask}
\end{figure*}

\noindent\textit{Uniform population density distribution}:
The methods under this category assume the population density is distributed homogeneously.
Figure~\ref{fig:cir_mask} illustrates the different methods under this category.  
\begin{enumerate}[left=0pt]
    \item \textit{Circular masking with a fixed radius}:
    This method generates random locations on a circle around the original 
    points by a predetermined displacement distance while ensuring that the 
    perturbed location points do not fall within the circle~\cite{kwan2004protection}.
    
    \item \textit{Circular masking with a random Radius}:    
    This method randomly displaces points both in direction and distance 
    within a circular area from their original positions, ensuring they 
    remain confined within the boundaries of the circle
    \cite{armstrong1999geographically,kwan2004protection}. All locations 
    inside the circle have an equal chance of being selected.
    
    \item \textit{Donut Masking (random perturbation within an annulus)}:
    The donut method 
    extends current methods of random displacement by ensuring that an address
    is not randomly assigned on or too near its original location. In donut 
    masking, each geocoded address is relocated in a random direction by at 
    least a minimum distance, but less than a maximum distance. 
    In population-based donut masking, each point is moved at a distance 
    inversely proportional to the underlying population density. This means 
    that points within regions of lower population density are displaced over 
    larger distances compared to those situated in areas of higher population 
    density. Essentially, the displacement magnitude is dynamically adjusted 
    based on the concentration of the population, ensuring that privacy 
    protection is maintained while minimising spatial error. Consequently, 
    points in sparsely populated regions undergo more substantial 
    repositioning compared to points in densely populated urban areas, thereby
    striking a balance between data privacy and spatial accuracy~\cite{stinchcomb2004procedures,allshouse2010geomasking}.
    
\end{enumerate}

\noindent\textit{Heterogeneous population density distribution}:
The methods under this category assume the population density is distributed 
non-homogeneously.

\begin{enumerate}[left=0pt]
    \item \textit{Spatially adaptive random perturbation (SARP)}:
    This technique considers the spatial variation of the population at risk 
    when displacing points by random distances and in random 
    directions~\cite{lu2012considering}. 
    It reduces unnecessary noise in high-dense population zones by using 
    relatively large and small population zones for low and high population
    densities, respectively. The SARP geomasking technique defines the 
    perturbation zone based on the actual distribution of residential 
    addresses (risk location) instead of people (risk population). 
    It employs "donut-shaped" spatially varied perturbation zones rather 
    than "pancake-shaped" which allows a point to be moved to a nearby 
    location within its immediate surroundings. Utilising actual street 
    addresses the SARP geomasking technique could offer better
    privacy protection compared to geomasking based on population size.

    \item \textit{Weighted random perturbation (WRP)}:
    This technique has the potential to preserve spatial distributions by displacing 
    points shorter distances in more populated areas~\cite{kwan2004}. The weighted 
    random mask aimlessly rotates every data point on a circle, applying the variable 
    radius $r$, where $r$ depends on the underlying population density.
    This approach imposes a maximum displacement limit based on the distance to each point's $k$-th nearest neighbour.
\end{enumerate}

\begin{figure*}[!t]
    \centering
    \includegraphics[width= 0.32\linewidth, keepaspectratio]{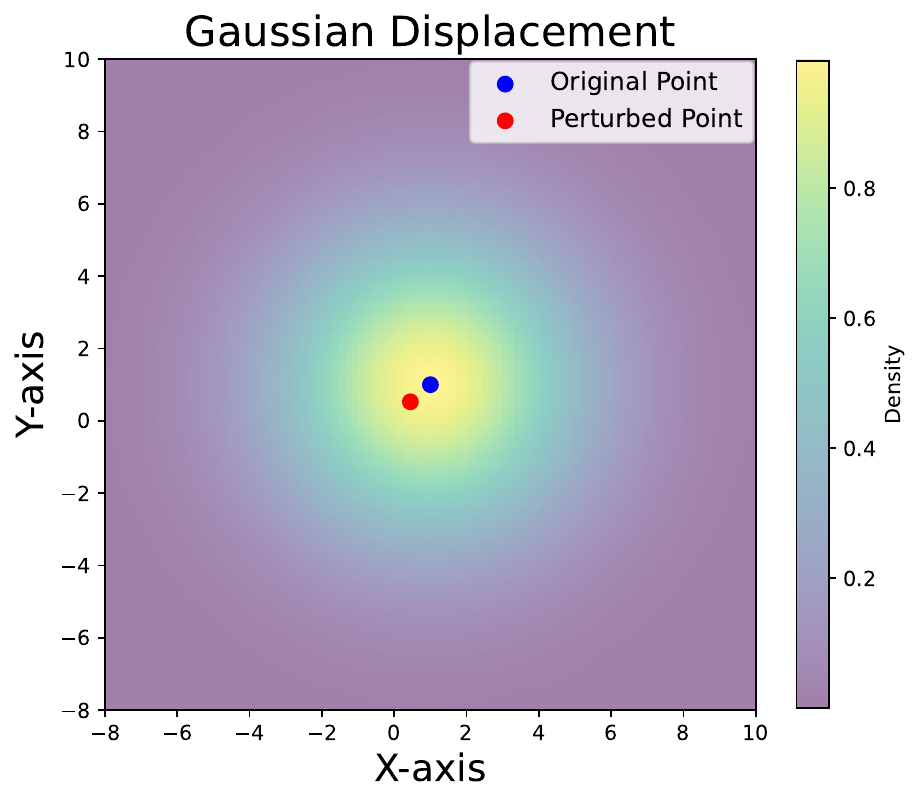}
    \includegraphics[width= 0.32\linewidth, keepaspectratio]{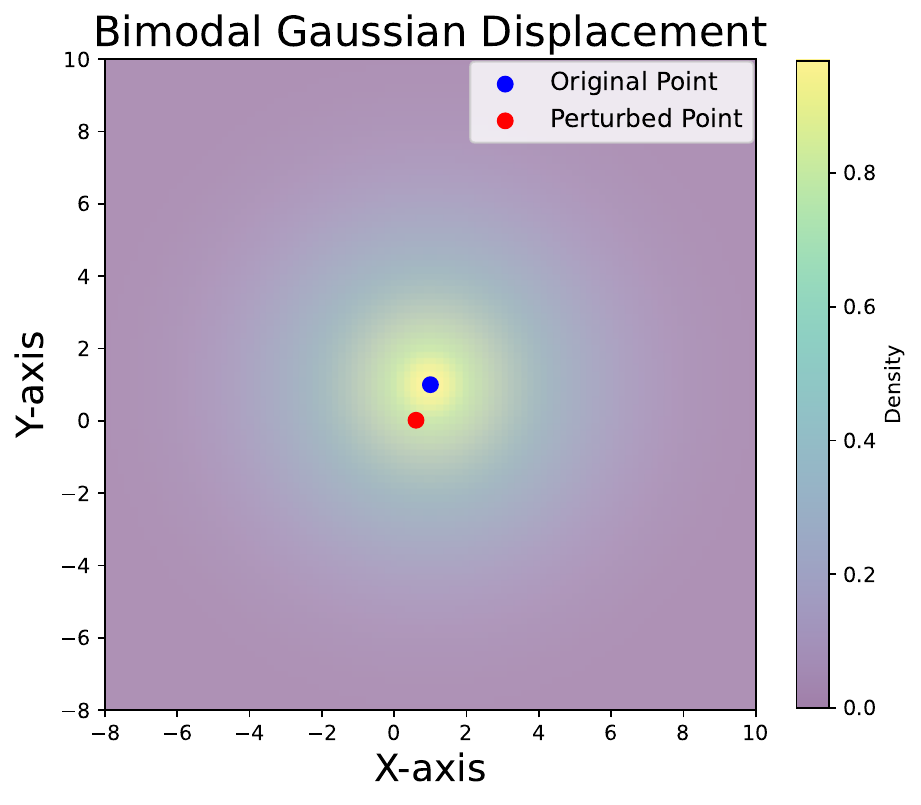}
    \includegraphics[width= 0.27\linewidth, keepaspectratio]{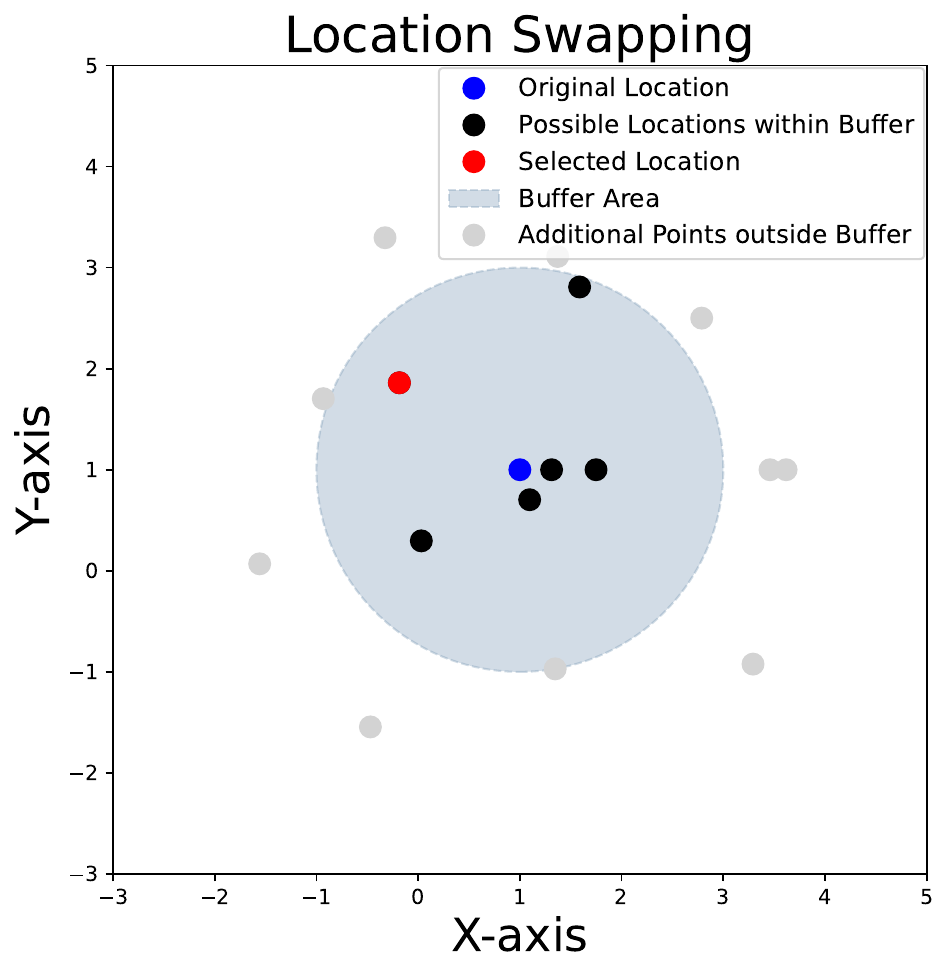}    
    \caption{Gaussian displacement (left), Bimodal Gaussian displacement (middle), and 
    location swapping (right).}
    \label{fig:gus_dist}
\end{figure*}

\noindent\paragraph{Gaussian deviation for displacement magnitude}
~\\
\noindent There are several methods that can be considered under this category. 
\begin{enumerate}[left=0pt]
    \item \textit{Gaussian displacement}:
    This method moves location points to a random direction,
    and the magnitude of displacement conforms to a Gaussian
    distribution~\cite{cassa2008re,fanshawe2011spatial,zandbergen2014ensuring}. 
    However, the Gaussian displacement’s main challenge lies in the limited 
    point displacement due to its concentration around the mean.
    Figure~\ref{fig:gus_dist} (left plot) illustrates this method. 
    
    \item \textit{Bimodal Gaussian displacement}:
    This technique is an adaptation of Gaussian displacement, employing a 
    bimodal Gaussian distribution to determine random distances for the masked 
    locations~\cite{cassa2006context,zandbergen2014ensuring}. It shares 
    similarities with donut masking but offers a probability
    distribution that is less uniform. Figure~\ref{fig:gus_dist} (middle plot) 
    illustrates this method.
    
    \item \textit{Density–based Gaussian spatial skew}:
    In this method, the location points are adjusted by applying a random offset
    derived from a Gaussian distribution, where the standard deviations of the 
    distribution are inversely related to the population density in the local
    area~\cite{cassa2006context}. Utilising local demographic data allows the 
    proposed anonymisation 
    system to shift individuals in heavily populated regions by shorter distances
    compared to those in sparse areas. Consequently, addresses can undergo 
    minimal distortion while upholding a designated level of k-anonymity.
    
\end{enumerate}

\noindent\paragraph{Multiple risk factor adaptive displacement}
~\\
\noindent This category incorporates various factors in calculating displacement distances,
including but not limited to population density. \textit{Triangular Displacement}
is a method that can be considered under this category. The triangular displacement method considers population density for displacement distances and incorporates multiple factors such as individual re-identification risk, 
data sensitivity levels, research types, the presence of quasi-identifiers,
and potential exposure to external sources to calculate displacement 
distances~\cite{murad2014protecting}. However, this technique has limitations.
It may be vulnerable to edge effects and could 
result in masked locations falling into impractical or unrealistic areas, 
thereby impacting data confidentiality. Despite these limitations, this 
method represents a step forward in geomasking techniques, offering a more 
nuanced and context-aware approach to safeguarding spatial data in research 
and analysis.

\subsubsection{Masking with pre-defined locations}
~\\
\noindent The proposed methods under this category move a given point to a new location, 
which is randomly chosen from all the possible locations of a similar type that 
exist within a specific region (i.e., a circle or an annulus) surrounding the 
original location. Location swapping methods offer advantages over 
random methods by providing a more realistic approach to selecting displaced locations, 
ensuring these locations fall within predefined areas of interest rather than arbitrary 
placements (i.e., bodies of water or uninhabited are). This enhances anonymity and 
preserves spatial patterns more akin to unmasked locations. However, if a malicious 
actor can ascertain the buffer distance used for relocation, there remains a risk 
of re-identifying the original locations~\cite{zhang2017location}.

\begin{enumerate}[left=0pt]
    \item \textit{Location Swapping}: This method swaps a given location/point to a new location based on the similar geographic characteristics of a point within a specified neighbourhood~\cite{zhang2017location}. The method first predefined a radius around the location to be swapped. Then, the method randomly selects a location among all the possible locations that fall within the defined radius. This method is similar to the random perturbation within a circle for a couple of reasons: 1) Within the predefined radius, all the locations are considered with equal probability, making the swapping to any location possible within the defined radius, and 2) the radius can be defined based on the underlying population density distributions. However, the location swapping method considered only the locations that have predefined locations (such as residential addresses) as possible candidates for swapping, and the radii can be varied based on the local population densities. Figure~\ref{fig:gus_dist} (right plot) illustrates an example of location swapping. 
    \item \textit{Location Swapping within Donut}: This method represents another variation of location swapping, employing the concept of donut masking to set a minimum distance, preventing points from being swapped with any locations within this specified range~\cite{zhang2017location}. The determination of this minimum distance can be based on local population densities. The authors employed an internal buffer size half that of the external buffer.
    \item \textit{Street Masking}: Similar to Location swapping, in the street masking method, the original location is replaced with coordinates of a point sampled from a set of points within a predefined radius~\cite{swanlund2020street}. Unlike location swapping, which relies on residential addresses, this method utilises road networks from a given map for sampling. During point selection, any points not in proximity to a drivable road (such as in rural areas) are further adjusted by moving them to the nearest applicable road. Similar to location swapping, street masking can leverage population densities to determine the radius, assuming areas with high population density also exhibit high road density and vice versa. Nonetheless, in areas with high road density, this method may experience longer runtimes due to the larger number of points to be masked.
\end{enumerate}
    Street masking offers a convenient solution for geographic masking by leveraging road network data, which is widely available globally. By aligning data points with street networks, street masking preserves topological continuity between original and masked locations and minimises false attribution risks, ensuring realistic shifts without arbitrary displacements. However, its reliance on road networks may introduce biases towards urban and built environments, potentially skewing analytical outcomes. In contrast, location swapping offers a viable alternative by allowing for more controlled displacement of data points without the inherent biases associated with street-centric masking. Location swapping is preferable in scenarios where avoiding point displacement onto roads is essential. This method overcomes issues related to urban biases and road network dependencies by swapping points with verified neighbors or alternative locations. Street masking lacks intuitive control over the search depth parameter, leading to varying masking distances influenced by local road configurations and density. This lack of control complicates the process of determining optimal settings, posing a challenge for users seeking consistent results across different datasets and study areas.

\begin{figure*}[!t]
    \centering
    \includegraphics[width= 0.3\linewidth, keepaspectratio]{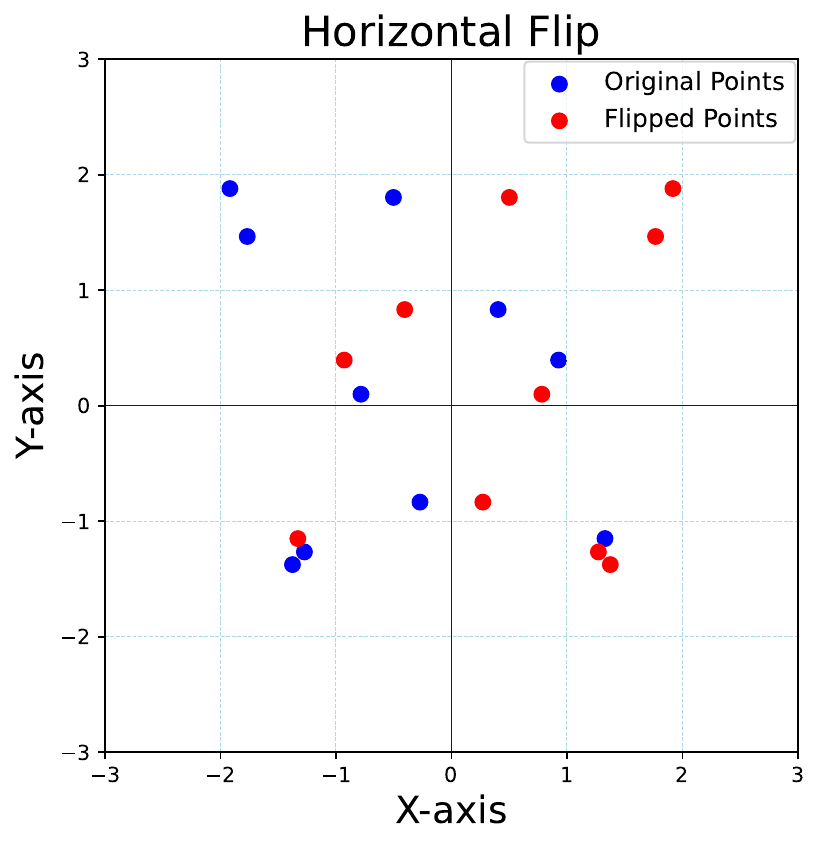}
    \includegraphics[width= 0.3\linewidth, keepaspectratio]{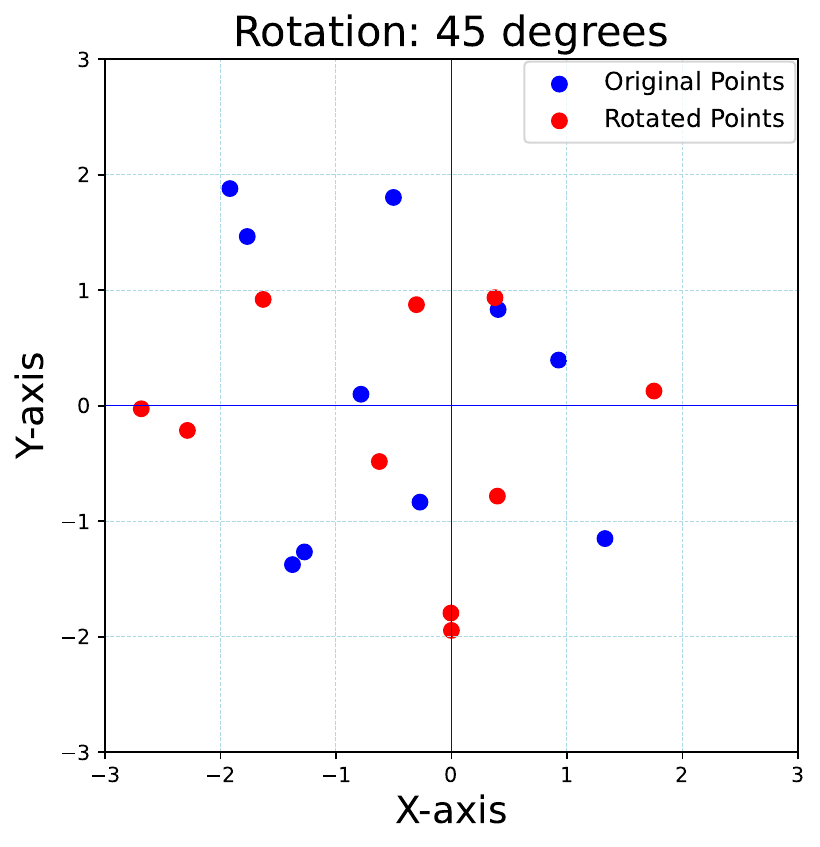}
    \includegraphics[width= 0.3\linewidth, keepaspectratio]{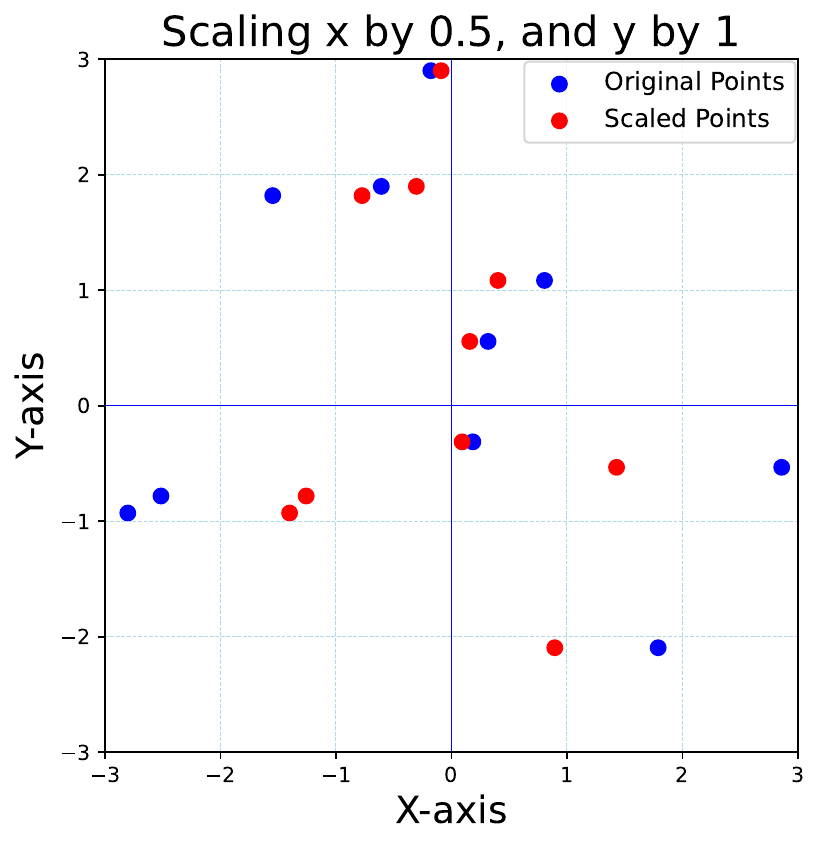}
    \caption{Affine transformation methods, Flipping (left), rotation (middle), 
    and scaling (right).}
    \label{fig:aff_trans}
\end{figure*}
\subsubsection{Affine-Transformation}
~\\
The transformation techniques, which are also known as isomasks techniques, apply 
basic geometric transformations to move spatial point data~\cite{ajayakumar2019addressing}. 
Here, it can be differentiated between rotate, scale and translate, stochastic 
mechanism, and concentration of isomasks. Affine transformations change the original location by either twisting every point by a set angle (referred to as rotation), moving every point by a set increment (referred to as translation) or moving every point by scaling constant (referred to as scale). Next, we describe these methods in more detail, and Fig.~\ref{fig:aff_trans} illustrates some of these methods.
\begin{enumerate}[left=0pt]
    \item \textit{Translation}: This method adds a random constant displacement value to both x and y coordinates
while ensuring the distance between coordinates remains the same~\cite{wang2022exploratory}. 
All points will be shifted to a fixed distance and direction.
    \item  \textit{Flipping}: This method manipulates the orientation of axes, both on a local and global scale. There are several flipping techniques, such as global horizontal flipping, which flips the map's horizontal axis; global vertical flipping, which flips the map's vertical axis; global axes flipping, which flips both horizontal and vertical axes; and local random flipping, which randomly alters the orientation of horizontal, vertical, or both axes within grid cells to hide location points \cite{gupta2020preserving,leitner2004cartographic,leitner2006first}.
    \item  \textit{Rotating}: This method rotates each point by a fixed angle ($0 \leq \theta \leq 360$), 
centering the rotation around either the origin of the coordinate system or an arbitrarily chosen point~\cite{wang2022exploratory}.
    \item  \textit{Scaling (Resizing)}: This method multiplies a constant value with both x and y coordinates, which remains consistent for both axes~\cite{wang2022exploratory}. Consequently, this process alters 
the distances between the points and all points will be expanded by a scaling factor.
    \item  \textit{Horizontal shear}: This method employs a linear transformation to horizontally shear the location points~\cite{armstrong1999geographically,broen2021measuring}. Each point is shifted along its x-axis, oriented at a 45° angle relative to its initial position concerning the center of the point distribution.
\end{enumerate}
Affine transformations preserve the spatial structure of the data, proving useful for spatial analysis and visualisation methods such as clustering, while maintaining the number of records and avoiding random relocation of data \cite{ajayakumar2019addressing}.  However, in these techniques, if the transformation matrix is known, retrieving the coordinates of the original records from the perturbed dataset is simple. Therefore, affine 
transformations are not a widely used methods~\cite{wang2022exploratory}. 

\subsubsection{Spatial aggregation \& Clustering}
~\\
\noindent 
Spatial aggregation and clustering techniques are commonly used to protect individual privacy by reducing spatial resolution, making it difficult to pinpoint exact locations and preventing the disclosure of precise geographical information. However, these techniques can also result in a loss of data granularity, which may restrict detailed analysis and reduce overall location accuracy.

\begin{figure*}[!t]
    \centering
    \includegraphics[width= 0.3\linewidth, keepaspectratio]{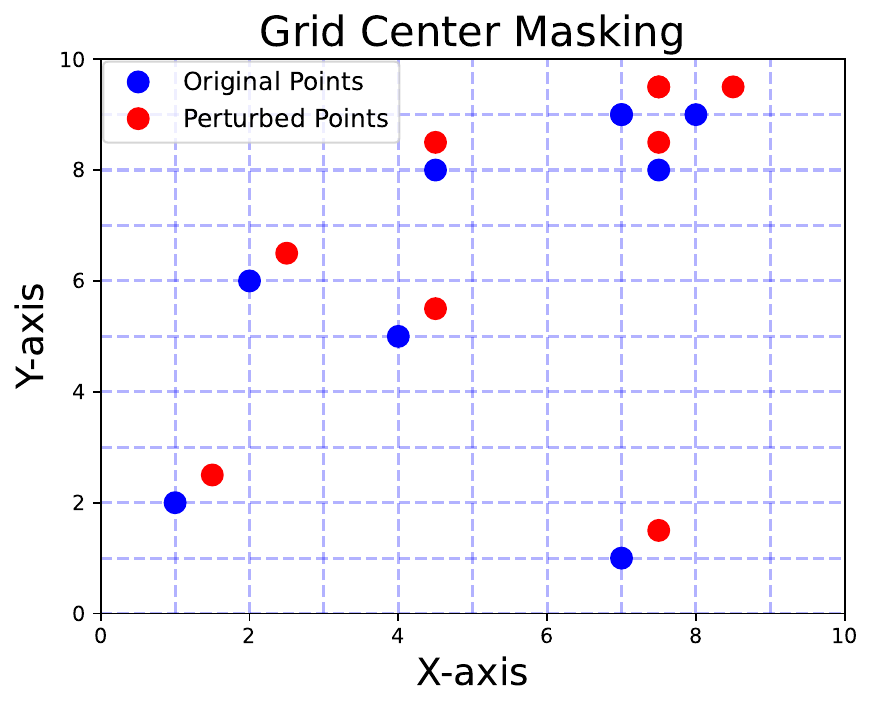}
    \includegraphics[width= 0.3\linewidth, keepaspectratio]{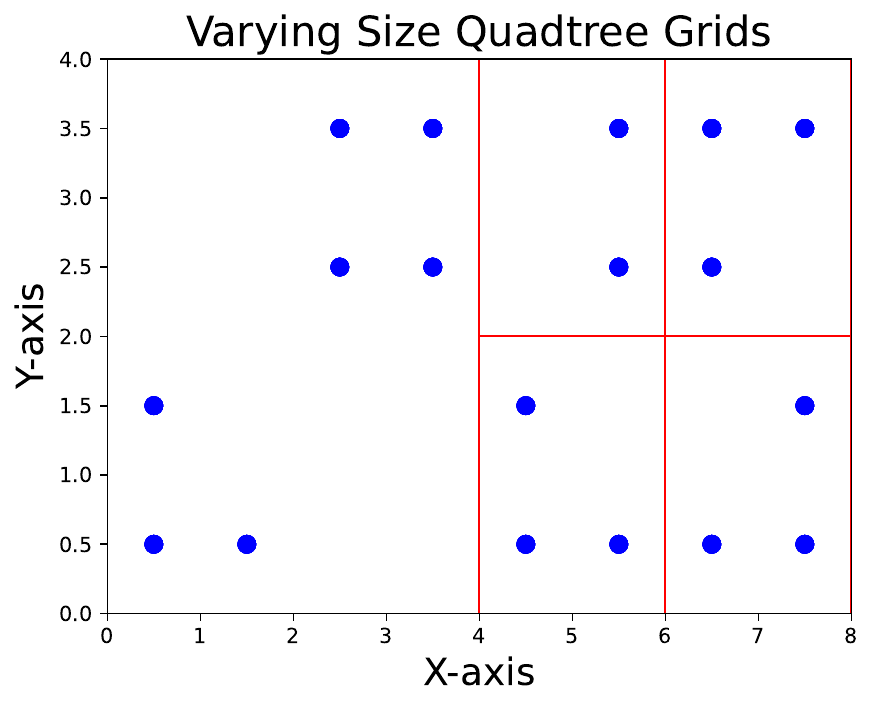}
    \includegraphics[width= 0.3\linewidth, keepaspectratio]{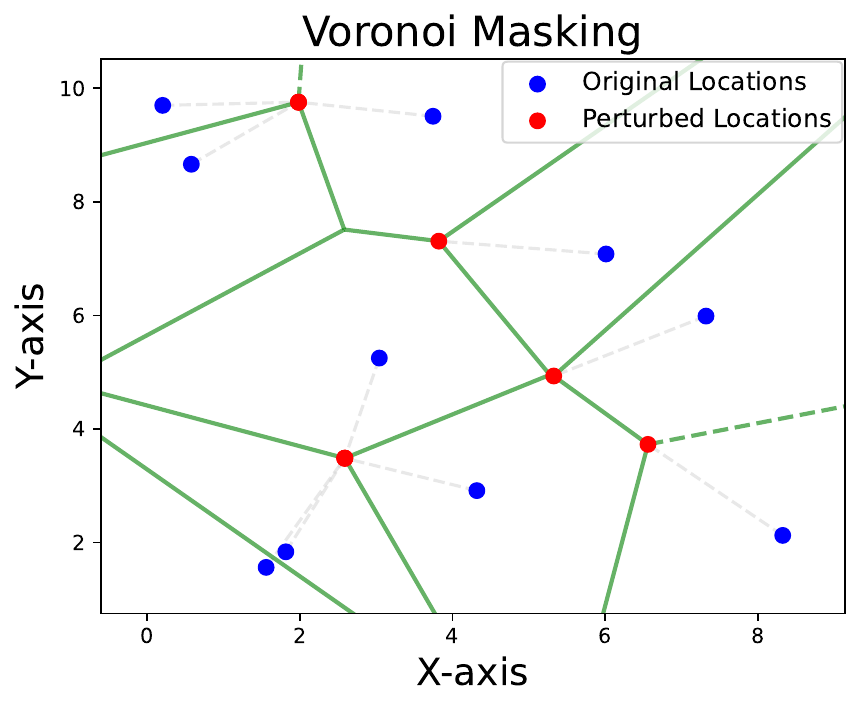}
    \caption{Grid center masking (left), quadtree grid (middle), and
    voronoi masking (right) techniques.}
    \label{fig:spa_agg}
\end{figure*}

Spatial aggregation aggregates individual location points within specific geographic
units or areas, such as cities, zipcodes, or administrative areas, to ensure anonymity
and prevent the disclosure of exact locations. This technique operates by decreasing
the spatial resolution of the data, thereby modifying the original data patterns. 
The goal is to generate generalised records that consolidate information within 
defined territories, protecting the identities of specific locations. However, 
this process may compromise the utility of published records as the detailed 
patterns found in high-resolution data might not be as discernible in the aggregated
information. Figure~\ref{fig:spa_agg} illustrates
different spatial aggregation techniques.

Clustering aims to reveal inherent patterns within spatial data by grouping similar 
data points based on geographical proximity or shared characteristics. 
Clustering maintains the overall characteristics of an area without pinpointing 
precise locations, preserving the general patterns while ensuring individual 
data privacy. However, due to the computations applied
directly on original data, clustering could expose
sensitive information especially in datasets containing personal or confidential
data. This can lead to unintended re-identification of individuals within the 
dataset, posing risks of privacy breaches and discrimination. Adversaries can 
exploit clustering results to deduce individuals' group memberships or glean 
details about the data distribution, even without direct access to the data~\cite{canbay2015effect}. 

\noindent\paragraph{Grid-based aggregations}
~\\
The techniques in this category involve the realignment of individual data 
points onto a grid system~\cite{leitner2004cartographic}. It typically involves 
relocating each original data point to specific positions within designated grid cells. 
These relocations may take place at either the centroid or vertices of the 
grid cells, each being aligned with a defined side length. The efficacy of methods
within this category is notably influenced by two crucial parameters: (1) the size of 
the aggregation units and (2) the boundaries utilised to define these units.

In grid-based aggregations, the grid cell size could be fixed or varied. 
\textit{Grid line masking} and \textit{grid center masking} are two popular methods that 
use fixed-size grid cells. The grid line masking method relocates location points
to the nearest edge of the grid cell in which they are 
contained~\cite{gupta2020preserving}. The grid center masking method relocates the 
location points to the centroid of the cell in which
they are presently situated~\cite{leitner2004cartographic}.
\begin{enumerate}[left=0pt]
    \item \textit{Quadtree hierarchical geographic data structures} is a method that uses
varying-size grid cells. This method employs a quadtree methodology on an initial
1km$^{2}$ European grid and population data points to create a grid with varying
sizes that can be adjusted according to local population 
density\cite{lagonigro2017quadtree}. In this approach, densely populated areas 
are divided into smaller square areas to ensure accurate data dissemination, 
while sparsely populated areas are grouped into larger areas to prevent the 
identification of individual data. 
\item \textit{Military Grid Reference System (MGRS)}
protects sensitive point-based geospatial data with a 
focus on preventing the matching of points with their true underlying 
observations, such as street addresses. This method involves converting 
coordinates to the Military Grid Reference System (MGRS) and using digit 
switching for masking, allowing encryption at different spatial precision 
levels \cite{clarke2016multiscale}. The selection of the digit switching 
combination minimises differences
in aggregate descriptive spatial statistics between the masked and 
unmasked data, with emphasis on the nearest neighbour statistic. 
Despite the large runtime, this method protect sensitive geospatial data, 
overcoming the disadvantages of geometric point masking methods, 
such as the lack of control over parameters, compromised spatial analysis 
of distribution, and susceptibility to violations of geoprivacy associated 
with the use of high-resolution imagery and accurate data from Global 
Navigation Satellite Systems (GNSS). The proposed digit-switching method, 
which operates effectively across multiple nested spatial scales, addresses
these limitations by allowing the users to choose a 
distribution that closely mimics the spatial distribution of unmasked data, 
ensuring suitability for spatial analysis based on aggregate descriptive 
statistics.
\end{enumerate}

\noindent\paragraph{Spatial tessellation-based techniques}
~\\
Spatial tessellation methods encompass techniques for dividing geographic 
areas into smaller, more manageable units while retaining the spatial 
integrity of the original data.

\begin{enumerate}[left=0pt]
    \item \textit{Voronoi masking}:
    Voronoi polygons define areas where the boundaries are equidistant between surrounding points or the interior of the polygons is closer to the corresponding point than to any other points~\cite{seidl2015spatial}. In Voronoi masking, each point
    is snapped to the closest part along the edges of its corresponding Voronoi polygon. 
    An advantage of this technique is that where the density of original points is 
    higher, the points are moved a shorter distance, resulting in patterns that 
    more closely resemble the original data. The Voronoi masking pattern is strongly linked to the distribution of points within the specified study area. Another 
    advantage of Voronoi masking is that some points in adjacent polygons will be 
    snapped to the same location, which can increase their k-anonymity.
    
    \item \textit{Adaptive areal elimination (AAE)}:
    This method establishes zones with a minimum level of 
    K-anonymity~\cite{kounadi2016adaptive}. 
    Initially, data points are grouped into polygons if available. Subsequently, 
    a disclosure value for "RoRi" (risk of re-identification) is determined, 
    representing the minimum K-anonymity threshold at which sensitive information
    can be revealed. The method involves a dissolving process, where polygons with
    RoRi values below the set threshold are merged with neighbouring polygons 
    until all polygons reach or exceed the threshold. This merging process 
    follows a spatial guideline based on shared borders, primarily prioritising
    neighbours with the longest shared border. In the case of equilateral 
    polygons, they undergo the dissolving process with all adjacent polygons. 
    This process results in the creation of K-anonymised areas 
    where the disclosure threshold is met. Location points within these areas
    can be relocated to the centroids or randomly
    distributed within the K-anonymised zones. 
    
    \item \textit{Adaptive areal masking (AAM)}:
    This method represents an enhanced iteration of 
    AAE~\cite{charleux2020true}. In AAM, specific 
    anonymisation polygons are constructed for individual points by iteratively
    merging them with the nearest centroid-based polygon, followed by the 
    next closest, until the k-anonymity threshold is met. This approach 
    preserves data more effectively than AAE because aggregation stops 
    when the desired k-anonymity level is reached. Additionally, 
    it minimises point displacement by initially merging smaller polygons. 
    Although merged polygons may not always be adjacent, this method is 
    more resource-efficient than verifying adjacency. It further 
    conserves resources by excluding polygons with no points to anonymise, 
    which can be a majority in some applications. Lastly, the process is 
    amenable to parallelisation as each point is treated individually, 
    enhancing efficiency.
    
    \item \textit{Adaptive voronoi masking (AVM)}:
    This method integrates the principles of AAE and VM to 
    preserve data utility, 
    considering topological polygon relationships and underlying topography,
    and achieve a high level of spatial k-anonymity to mitigate 
    re-identification risks\cite{polzin2020adaptive}. A key strength of AAE 
    lies in accounting for underlying population density, while 
    VM excels in displacing the 
    Original Data Points (ODP), generating masked patterns similar to 
    the ODP's distribution. However, both VM and AAE exhibit drawbacks: 
    they overlook the underlying topography, potentially placing data 
    points in illogical locations such as lakes or forests, leading to 
    false re-identification risks. AVM bridges these gaps by integrating
    the benefits of AAE's population density consideration and VM's ODP 
    displacement method. It moves ODPs to the closest segments of their 
    corresponding Voronoi polygon within the merged AAE-polygon, 
    mitigating the risk of displacing points to areas with different 
    population thresholds. Additionally, AVM addresses topographical 
    concerns by relocating data points to the nearest street intersection,
    minimising false re-identification and illogical placements.
    
    \item \textit{RDV masking}:
    This method, known as the three-layer RDV masking model, is based on 
    rudimentary triangulation and Voronoi generation\cite{gupta2020preserving}. 
    The masking process is centralised, applying masking to specified nodes 
    at one point. The key advantage of this solution is its ability to not 
    only efficiently mask points but also retain the spatial pattern of the 
    original data, considering various indicators beyond population density. 
    The RDV masking method is structured into three distinctive layers, each 
    contributing to the overall masking process. The first layer, termed layer R,
    initiates by spatially distributing a given set of points throughout 
    the region, ensuring the preservation of the original point configuration
    for later comparison and analysis. Moving to layer D, the process 
    progresses to computing the Delaunay triangulation of the given point 
    set, utilising a Divide and Conquer strategy to efficiently accomplish 
    this phase, albeit with a specific computation duration. Finally, layer 
    V, the last phase, focuses on separating individual points to their 
    nearest segment along the edges of the corresponding Voronoi polygon. 
    This step is pivotal in upholding the original density patterns of 
    the points, employing Fortune’s Algorithm for computation, which 
    also entails its specific computation duration. Together, these layers
    encapsulate a comprehensive approach to masking, offering spatial 
    distribution, triangulation, and precise segmentation to maintain 
    original data patterns effectively.
    \item \textit{Zip4codes aggregation}: This method uses Zip4 (postal) 
    codes as a spatial aggregation unit for sharing fine-scale health 
    data while addressing privacy concerns~\cite{ajayakumar2023utility}. 
    The Zip4 aggregation preserves the underlying spatial structure, making
    it potentially suitable for geospatial analysis.
    
\end{enumerate}
\noindent\paragraph{Differentially private clustering}
~\\
\noindent Differential privacy (DP), recognised as the current 
state-of-the-art provable privacy technique for mitigating the risk of 
identity disclosure by incorporating randomisation and noise addition~\cite{Dwo06}. 
By leveraging randomisation techniques coupled with noise addition, DP significantly
diminishes the disclosure risk associated with these methods, offering robust 
protection for sensitive data. An algorithm is said to be differentially private if 
the algorithm produces essentially the same answers
regardless of the presence or absence of an individual in the dataset.

\begin{figure*}[!t]
    \centering
    \includegraphics[width= 0.3\linewidth, keepaspectratio]{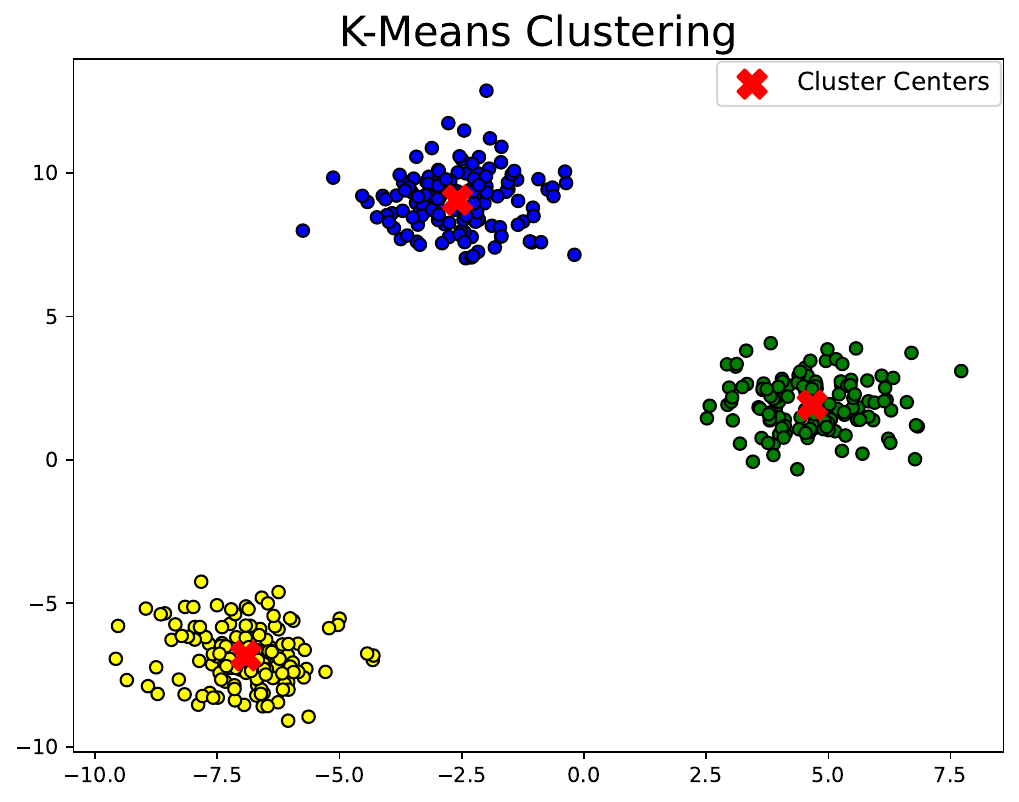}
    \includegraphics[width= 0.3\linewidth, keepaspectratio]{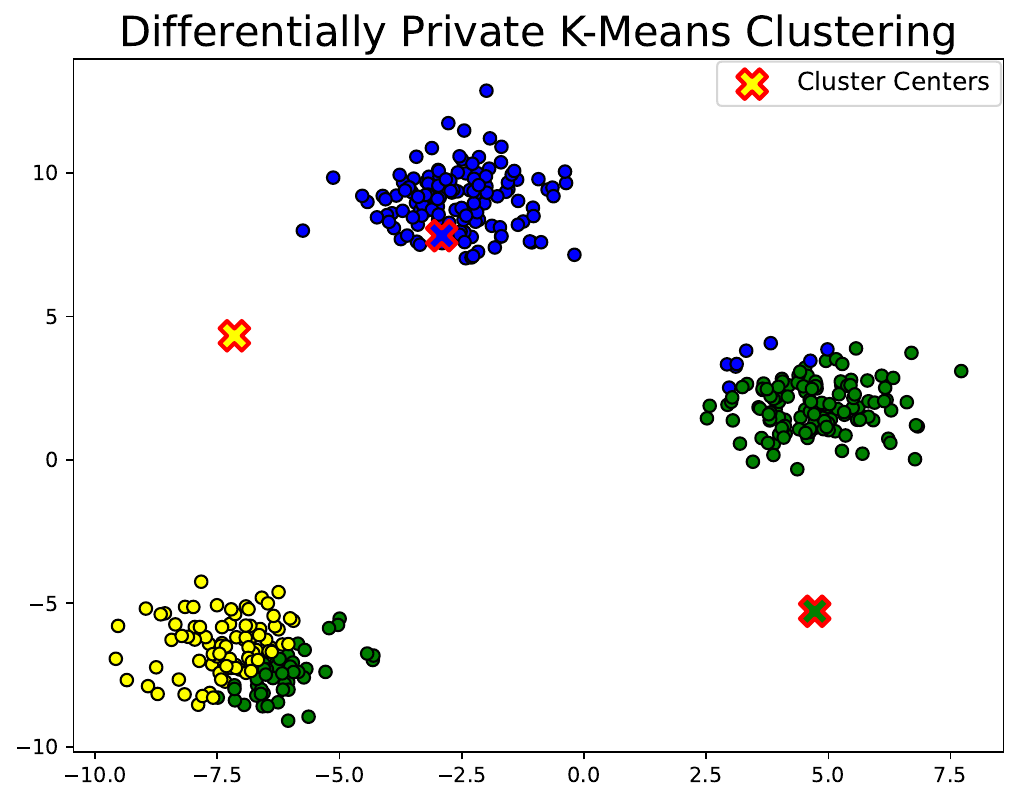}
    \caption{K-means clustering (left) and differentialy private k-means clustering (right).}
    \label{fig:dp_clustering}
\end{figure*}

DP has recently been used for querying geospatial datasets in a 
privacy-preserving manner~\cite{xiao2015, harris2020}. In addition to its application
in statistical queries, DP has been employed in geospatial clustering. 
Clustering geospatial data can inadvertently expose vulnerabilities. For example, 
the disclosure of cluster centroids during the iterative clustering 
process can lead to the potential leakage of information about individual data 
points, particularly when adversaries have background knowledge or auxiliary 
information enabling them to conduct inference attacks~\cite{lu2020differentially}.

DP techniques could inject noise into the clustering process such that the 
centroids may move slightly or significantly from their original positions 
depending on the magnitude and distribution of the injected 
noise~\cite{yan2022, wang2017, qardaji2013,lu2020differentially}.
Figure~\ref{fig:dp_clustering} 
illustrates both the original k-means clustering (on the left) and the 
differential privacy-based k-means clustering (on the right) applied to the
same dataset. To mitigate such risks and safeguard individual privacy, 
considerable research effort has been dedicated to integrating differential 
privacy into clustering algorithms.

The main drawback of using DP algorithms for data analysis is that, like other perturbation methods, they do not preserve the accuracy of the algorithms applied to the original data. In fact, they tend to deviate more from the results based on the original data in proportion to the privacy guarantees they provide~\cite{balcan2017differentially}. 
Further, DP 
implicitly assumes that all information about an individual participant is 
contained within a record and that records are independent. Thus, if the dataset 
contains multiple location records per person the the DP mechanism needs to be 
design to capture all these records of an individual. If location records are 
correlated, the privacy protections may weaken, as the DP mechanism may 
inadvertently leak information due to the interdependence of records.

\subsubsection{Synthetic data generation}
~\\
\noindent Synthetic datasets
aim to preserve the spatial characteristics and patterns present in the original
data while preserving the individual's privacy. Synthetic data can be generated in a number 
of ways~\cite{bauer2024}. 
Some synthetic data generation techniques include (1) Generative AI; these 
techniques are based on Generative Pre-trained Transformers (GPT), Generative 
Adversarial Networks (GANs), or Variational Auto-Encoders (VAEs) which learn
the underlying distribution of real data to generate similarly distributed 
synthetic data~\cite{zhao2021, wu2022}. (2) A rules engine that can create 
synthetic data via user-defined business rules. Intelligence can be added to 
the generated data by referencing the relationships between the data elements, to ensure the 
relational integrity of the generated data~\cite{mckenzie2023}. (3) Entity 
cloning is another technique that extracts data from the source systems 
of a single entity (e.g., patient) and masks it for compliance. It then 
clones the entity, generating different identifiers for each clone to 
ensure uniqueness. (4) Data masking is also a possible option that replaces personally identifiable information with fictitious, yet structurally 
consistent, values, such as addresses or coordinates~\cite{nowok2018}. 
The objective of data masking is to ensure that sensitive data cannot be 
linked to individuals, while retaining the overall relationships and 
statistical characteristics of the data. 

Some other techniques, such as Copula models, could discover correlations
and dependencies within the datasets, and then use these correlations and 
dependencies to generate new, realistic data. Data augmentation (such as flipping,
rotation, scaling, and translation) could also be applied as supplementary techniques to existing values to create new data. Finally, Techniques such as random sampling and noise injection, which add data points from known distributions, could also be used to create new data points that closely resemble 
real-world data~\cite{sakshaug2014}.

\subsection{Evaluation measure for Geomasking}
The performance of the geomasking techniques can be evaluated in terms of how 
efficient and effective they are when applied to geospatial datasets. The efficiency 
of a geomasking technique evaluates how scalable a masking technique is on a large 
dataset with millions of records, while the effectiveness of a geomasking technique
is measured by the accuracy and the protection it can provide for an individual. Next, 
we discuss different measures that have been proposed to evaluate geomasking 
techniques in terms of scalability, accuracy, and privacy.

\subsubsection{Scalability}
~\\
\noindent Scalability can be evaluated using measures that are dependent on the computing
platform and networking infrastructure used, or measures that are based on the
number of data points analysed~\cite{redlichquantitative}. Common measures include
run time, memory space, and communication size. 

Apart from these measures, the scalability can also be evaluated theoretically 
using computational complexity measures. This includes the computation and 
communication complexities (if multiple data providers are available) that measure
the overall computational efforts and cost of communication required in the 
geomasking process. Generally, the big O-notation is used to specify the 
computation complexity~\cite{papadimitriou2003}. Given $n$ is the number of records 
in a database, the big O-notation of represents $O(log\ n)$ logarithmic complexity, 
$O(n)$ linear complexity, $O(n\cdot log\ n)$ log-linear complexity, $O(n^2)$ 
quadratic complexity, $O(n^c)$ polynomial complexity, $O(polylog\ n)$ polynomial 
logarithmic complexity, and exponential complexity, where $c>1$.

\subsubsection{Accuracy}
~\\
\noindent Table \ref{tab:metric} presents our classification of metrics for evaluating the accuracy of geomasking techniques. These metrics are categorised into three groups: Spatial Point Pattern \& Distribution Analysis, Displacement Distance Analysis, and Spatial Autocorrelation Analysis. 
\begin{table}[!ht]
\caption{Metrics for evaluating the accuracy of geomasking techniques}
\label{tab:metric}
\begin{footnotesize}
\begin{tabular}{|l|l|}
\hline
\textbf{Accuracy Metric Category} & \textbf{Metric} \\ \hline
\multirow{13}{*}{Spatial Point Pattern \& Distribution Analysis} & Average Nearest Neighbor (ANN) \\
 & Kernel Density Estimation (KDE) \\
 & Point density estimation (PDE) \\
 & Ripley’s K \\
 & L-function \\
 & Cross-K function \\
 & Minimum convex polygon (MCP) \\
 & Standard deviation ellipse (SDE) \\
 & Global divergence index (GDi) \\
 & Local divergence index (LDi) \\
 & Spatial cluster detection \\
 & Spatial central tendency and dispersion (normal distribution) \\
 & Spatial central tendency and dispersion (non-normal distribution) \\ \hline
\multirow{2}{*}{Displacement Distance Analysis} & Displacement distance \\
 & Displacement statistical summary \\ \hline
\multirow{2}{*}{Spatial Autocorrelation Analysis} & Global Moran’s I \\
 & Local Moran’s I \\ \hline
\end{tabular}%
\end{footnotesize}
\end{table}
\paragraph{Spatial Point Pattern \& Distribution Analysis}
~\\
This category assesses points' spatial arrangement and distribution in both the original and masked datasets. 
\begin{enumerate}[left=0pt]
    \item \textit{Average Nearest Neighbour (ANN)}: 
    ANN is a spatial analysis method used to quantify the arrangement of points within 
    a dataset\cite{iyer2023advancing, wang2022exploratory, ajayakumar2023utility,broen2021measuring,seidl2015spatial,zandbergen2014ensuring}. 
    It calculates the average distance of each point to its nearest neighbour, offering
    valuable insights into the spatial pattern of the data. This metric is particularly
    useful for identifying clustering at a global level, providing valuable information
    about the spatial distribution and clustering tendencies of the dataset.
    
   \item \textit{Kernel Density Estimation (KDE)}:
   KDE is a robust point pattern analysis technique designed to create smooth density
   surfaces from point feature datasets\cite{iyer2023advancing, wang2022exploratory, seidl2018privacy, ajayakumar2023utility,zandbergen2014ensuring}.  
   It is particularly useful for detecting clusters and evaluating the impact of 
   displacement, as seen in geomasking. KDE allows for the quantitative assessment of 
   spatial clustering. This comparison not only aids in identifying location tendencies
   but also provides an estimation of the extent to which the original spatial 
   patterns are preserved after masking.
   
   \item \textit{Point density estimation (PDE)}:   
   PDE entails constructing density surfaces or maps from point feature datasets
   to visually depict spatial patterns~\cite{wang2022exploratory}.
   
   \item \textit{Ripley's K}:    
    This metric analyses spatial point process data, revealing clustering patterns 
    at various local distances~\cite{dixon2001ripley,ajayakumar2023utility,swanlund2020street}.
    It summarises point patterns across multiple scales, allowing the determination of clustering occurrences at different distances and underlying spatial patterns.
    
    \item \textit{L-function}:    
    This metric is a normalised version of Ripley's K that evaluates the expected
    count of points within a multi-dimensional sphere with a radius $r$, 
    normalised by the volume of the sphere~\cite{dixon2001ripley,broen2021measuring}. 
    It determines whether the observed distribution of points in close proximity to a particular location indicates either clustering or dispersion, beyond what would be anticipated 
    under random chance.
    
    \item \textit{Cross-K function}:    
    This metric assesses the spatial clustering patterns between two distinct point patterns~\cite{kwan2004protection,seidl2015spatial,zandbergen2014ensuring}. 
    The Cross-K function quantifies the expected number of points from the reference
    pattern ($j$) within a specified distance of an arbitrary point in the target 
    pattern ($j'$), normalised by the overall density of points within the 
    reference pattern ($j$). This measure explores relationships among multiple
    point patterns in spatial statistics.
    
    \item \textit{Minimum convex polygon (MCP)}:    
    MCP is a spatial analysis method used to define the minimum convex hull that 
    encompasses a set of points or features within a dataset, presenting the 
    spatial arrangement of features within a dataset~\cite{wang2022exploratory}.
    
    \item \textit{Standard deviation ellipse (SDE)}:    
    SDE is utilised to visually represent the spatial distribution of features 
    in a dataset~\cite{yuill1971standard,wang2022exploratory, seidl2018privacy}. 
    It serves as a tool for measuring the spread or variability of 
    a point pattern and assessing the concentration of geographic coordinates.
    
    \item \textit{Global divergence index (GDi)}:   
    GDi is specifically designed for summarising disparities between spatial 
    data~\cite{kounadi2015spatial, seidl2018privacy, nowbakht2022comparison}. 
    GDi quantifies the dissimilarity of masked data in relation to the original data, 
    characterising the spatial information divergence by considering three essential 
    components: the divergence of spatial mean, the orientation, and the major axis of the first standard deviational ellipse. GDi produces an average divergence index that effectively represents the dissimilarity of the obfuscated point pattern to its original counterpart. 
    
    \item \textit{Local divergence index (LDi)}:    
    LDi is based on the symmetrical difference between nearest neighbor
    hierarchical clustering and Getis-Ord Gi* hot spots in comparison to the 
    original point distribution, providing a means for concise point distribution 
    comparisons~\cite{kounadi2015spatial,seidl2018privacy, nowbakht2022comparison}. 
    It calculates divergence by evaluating the disparity in hotspot areas between 
    the obfuscated and original points.
    
    \item \textit{Spatial cluster detection}:    
    This metric evaluates the presence of clusters within a set of spatial data 
    points~\cite{seidl2018privacy, ajayakumar2023utility}. Commonly used spatial 
    clustering tools include the Cuzick-Edwards statistic \cite{cuzick1990spatial}, 
    SaTScan circular clustering detection \cite{kulldorff2006information}, and 
    GeoMEDD\cite{curtis2020geographic}.
    
    \item \textit{Spatial central tendency and dispersion (normal distribution)}:    
    Metrics such as minimum (min), maximum (max), mean, mode, median, standard 
    deviation (SD), and Standard Deviational Ellipse are fundamental statistical 
    measures offer insights into the impact of geomasking methods on spatial distortion. 
    Calculated for both the original and masked datasets, these measures enable a 
    comprehensive understanding of how the central tendency and variability of 
    spatial data are influenced~\cite{redlichquantitative,nowbakht2022comparison}. 
    
    \item \textit{Spatial central tendency and dispersion (non-normal distribution)}:    
    Measures such as min, max, median, 1st quartile, 3rd quartile, and the interquartile range (IQR) are useful for characterising the central tendency 
    and dispersion of point data within a dataset~\cite{nowbakht2022comparison}. 
    These metrics allow for a detailed assessment of distortion in datasets with 
    non-normal distributions, contributing significantly to refining geomasking 
    methodologies and interpreting spatial data in various contexts.
\end{enumerate}

\paragraph{Displacement Distance Analysis}
~\\
This category measures the distance of displacement between original and masked locations.
\begin{enumerate}[left=0pt]
    \item \textit{Displacement distance}:    
    This metric is employed to measure the distance between the masked location and 
    its original location~\cite{swanlund2020street,redlichquantitative}.  It provides
    insights into the extent of spatial distortion introduced by the masking process. 
    
    \item \textit{Displacement statistical summary}:    
    This metric is utilised to quantify the number or percentage of points
    within a dataset exhibiting distinct displacements~\cite{nowbakht2022comparison}. 
\end{enumerate}

\paragraph{Spatial Autocorrelation Analysis}
~\\
This category examines the spatial correlation of both original and masked data points.
\begin{enumerate}[left=0pt]
    \item \textit{Global Moran’s I}:    
    This metric is used to assess global-level spatial autocorrelation, providing 
    insights into the distribution pattern of a variable across 
    geographic space~\cite{moran1950notes,broen2021measuring,redlichquantitative,zandbergen2014ensuring}.
    Ranging from -1 (indicating complete separation) to 1 (reflecting complete clustering), 
    this index serves as a valuable tool for quantifying the degree of spatial
    clustering or dispersion in a dataset. 
    
    \item \textit{Local Moran’s I}:    
    This metric is used to assess the presence of local spatial autocorrelation within 
    a geographic dataset~\cite{anselin1995local,broen2021measuring}. In contrast to the 
    Global Moran's I, Local Moran's I offers a finer-grained analysis by examining the 
    similarity between a specific spatial unit and its neighbouring areas. 
    This metric is particularly valuable for uncovering patterns of spatial 
    autocorrelation that may be concentrated in specific regions rather than uniformly
    spread across the entire geographic area. By identifying hotspots and cold spots,
    Local Moran's I aids in the identification of localised trends, clusters, or 
    outliers, contributing to a more nuanced understanding of spatial relationships 
    within the dataset.
\end{enumerate}

\subsubsection{Privacy} \label{sec:privacy}
~\\
In this section, we discuss several metrics that can be used to measure the privacy of the geomasking techniques. 

\begin{enumerate}[left=0pt]
    \item \textit{Spatial k-anonymity}: 
    Spatial k-anonymity is a crucial metric in assessing the disclosure risk of 
    geomasked datasets~\cite{allshouse2010geomasking,seidl2015spatial,iyer2023advancing, wang2022exploratory, seidl2018privacy, dupregeospatial, broen2021measuring,redlichquantitative, ajayakumar2023utility,zandbergen2014ensuring}.
    This method evaluates the likelihood of identifying a specific record within a 
    given spatial region by calculating the number of potential geospatial records 
    in proximity to that record. Spatial k-anonymity 
    leverages reverse geocoding as a disclosure threat mechanism. 

    One approach to quantifying spatial k-anonymity entails examining the number or
    percentage of obfuscated points within a 
    specified distance, considering varying values of $k$. In the geomasking context, 
    the anonymisation region is created as a buffer with a radius ($r$) centred at the
    masked location, where $r$ corresponds to the distance between the original and 
    masked location. A higher k value reduces the probability of successfully
    re-identifying the original location, indicative of enhanced geomasking performance.
    Spatial k-anonymity can also be determined by counting the number of location 
    points within a radius ($r$) for each specific point, where $r$ represents 
    the maximum displacement distance resulting from the masking approach. 
    This calculation is applicable when considering the weight of displacement
    in masking and evaluating the level of privacy achieved after masking. 
    
    Spatial k-anonymity is often determined using the $n$-th nearest neighbour 
    calculation, referring to the number of potential locations closer to the masked
    location than to the original location. This approach essentially reveals the 
    empirical estimate of spatial k-anonymity, providing insights into the 
    reidentification risk associated with masked datasets.  Additionally, spatial 
    k-anonymity can be measured as the average distance to k nearest neighbors for
    the masked point. A lower distance to k neighbors suggests a lower probability of
    association with the original location, indicating a higher level of privacy.

    \item \textit{Topological identification risk}:    
    Apart from spatial k-anonymity, this measure considers the topological information in 
    masked data~\cite{seidl2018privacy}. The metric assesses the privacy risk in geomasking
    through the identification of topological relations between masked points. This measure conceptualises false identification risk considering factors such as topological relations,
    the existence of a single nearest neighbour location, and the density of neighbouring
    locations. The identification risk could be categorised into various scenarios, including 
    points interior or disjoint to correct or incorrect locations, on the boundary of correct
    and incorrect locations, and disjoint to locations with correct or incorrect nearest
    neighbours. 
\end{enumerate}

Apart from the above measures, Redlich~\cite{redlichquantitative} highlights 
additional re-identification risk assessments. These measures include linking the 
geodata with a dataset containing longitudinal information, mean of multiple releases,
and reversing masking techniques. 
Linking geodata with longitudinal information involves mapping masked geospatial 
records with a dataset containing direct identifiers about individuals, such as 
names and addresses.   

For the linking process, two attack schemes have been proposed. (1) Graph theoretic
linkage attack~\cite{kroll2015} uses graph theory to link masked and original 
coordinates, framing the problem as a maximum clique issue. (2) Graph matching 
attack~\cite{vidanage2020} uses graph matching mechanisms to identify matching 
records in encoded and plaintext datasets. The attack utilises features like 
node-based, edge-based, and structural features, applying cosine locality-sensitive 
hashing for identifying matching records. 
The second measure, mean of multiple releases, involves using various dataset 
versions to determine the original coordinates' location. The idea is based on the 
central limit theorem, which states that as the number of observations increases, 
the average of the masked coordinates will fall closer to the original point. The 
third measure, reversing masking techniques, aims to reverse a masking method, such as affine 
transformation, by testing all possible parameters to find the original location.

\subsection{Practical aspects of Geomasking}

The final three dimensions cover practical aspects of geomasking techniques, 
including the datasets used for experimental evaluations, how a solution was 
implemented, and various application domains where geomasking can play a pivotal role.

\subsubsection{Datasets}
~\\
To evaluate the practical
aspects of geomasking techniques with regard to their expected performance 
in real-world applications, evaluations should ideally be done on databases
that exhibit real-world properties and error characteristics. However, in
some occasions, due to the difficulties of obtaining real-world data that 
contain geospatial information, synthetically generated geo databases could 
also be used.
Table~\ref{tab:datasets} provides details of several datasets
that have been used by different studies for experiments with geomasking 
techniques. It is important to note that though these datasets are available publicly, 
some of these datasets might have ethical obligations. 

\begin{footnotesize}
\begin{longtable}{|>{\raggedright\arraybackslash}p{1.8cm}|>{\raggedright\arraybackslash}p{7.3cm}|p{2.3cm}|>{\centering\arraybackslash}p{1cm}|}

\caption{Datasets used by different studies for experiments on geomasking techniques.} \label{tab:datasets} \\
\hline
\textbf{Dataset} & \textbf{Description} & \textbf{Repository Link} & \textbf{Used by} \\ \hline
\endfirsthead

\multicolumn{4}{c}%
{{\tablename\ \thetable{} -- Continued from previous page}} \\
\hline
\textbf{Dataset} & \textbf{Description} & \textbf{Repository Link} & \textbf{Used by} \\ \hline
\endhead

\endfoot

\hline
\endlastfoot

 European data & The official portal for European data & \scriptsize\url{https://data.europa.eu/data/datasets/?locale=en} & \cite{polzin2020adaptive} \\ \hline
 French public data & Open platform for French public data.
 \cite{houfaf2021geographically} utilised below datasets
 \begin{itemize}[left=0pt,itemsep=0pt,parsep=0pt,topsep=0pt,partopsep=0pt]
     \item [-]	\footnotesize COVID-19 virological test results on\newline \url{https://www.data.gouv.fr/fr/datasets/old-donnees-relatives-aux-resultats-des-tests-virologiques-covid-19/} 
    \item [-] \footnotesize Base National Address on \url{https://www.data.gouv.fr/fr/datasets/base-adresse-nationale/}
 \end{itemize}& \scriptsize\url{https://www.data.gouv.fr/fr/datasets/} & \cite{houfaf2021geographically} \\ \hline
US Census Bureau data & This repository provides data related to People, Places, and Economy; covering over 100,000 different geographies: States, counties, places, tribal areas, zip codes, and congressional districts in the United States. For each, they cover topics including, but not limited to, education, employment, health, and housing.
\begin{itemize}[left=0pt]
    \item[-] \footnotesize\cite{wang2022exploratory} utilised United States Census Bureau, 2021
    \item[-] \footnotesize\cite{dupregeospatial} utilised United States Census Bureau, 2019 (Income and Poverty in the United States: 2019)
\end{itemize}
 & \scriptsize\url{https://data.census.gov/} & \cite{wang2022exploratory, dupregeospatial} \\ \hline
 US GIS data & The NYS GIS Clearinghouse is an evolving searchable repository of GIS data and mapping resources. \cite{zhang2017location} utilised actual residential datasets from three U.S. counties in their analysis, specifically examining data from Jackson County in Oregon, Travis County in Texas, and Wake County in North Carolina. & \scriptsize\url{https://data.gis.ny.gov/} & \cite{zhang2017location} \\ \hline
 U.S. Census Bureau's geographic spatial data & This repository represents the U.S. Census Bureau’s geographic spatial data, which has been developed using Topologically Integrated Geographic Encoding and Referencing system (TIGER) geospatial data as the primary source.
 & \scriptsize\url{https://www.census.gov/programs-surveys/geography/guidance/tiger-data-products-guide.html} & \cite{su2017differentially} \\ \hline
  US GIS data & The NYS GIS Clearinghouse is an evolving searchable repository of GIS data and mapping resources. \cite{zhang2017location} utilised actual residential datasets from three U.S. counties in their analysis, specifically examining data from Jackson County in Oregon, Travis County in Texas, and Wake County in North Carolina. & \scriptsize\url{https://data.gis.ny.gov/} & \cite{zhang2017location} \\ \hline
  U.S. Department of Transportation (USDOT) & the National Address Database (NAD), established by the USDOT and its partners, provides addresses aggregated from state, local, and tribal governments. These addresses are compiled to provide accurate and up-to-date information critical for various purposes, including transportation safety, Next Generation 9-1-1 services, mail delivery, permitting, and school siting.
 & \scriptsize\url{https://www.transportation.gov/gis/national-address-database} & \cite{lin2023geo} \\ \hline
 Australian Government open data & Data.gov.au serves as the primary repository for open government data in Australia. Individuals can retrieve anonymised public data released by federal, state, and local government agencies. & \scriptsize\url{https://data.gov.au/search} & \cite{redlichquantitative} \\ \hline
 Geoscape Geocoded National Address File (G-NAF) & Geoscape G-NAF is a trusted source of geocoded address database for both Australian businesses and governmental entities, comprising more than 50 million submitted addresses condensed into 15.4 million G-NAF addresses. It
is built and maintained by Geoscape Australia using independently examined and validated government data. \cite{redlichquantitative} utilised August 2019 - PSMA Geocoded National Address File (G-NAF) from \scriptsize\url{https://data.gov.au/dataset/ds-dga-e1a365fc-52f5-4798-8f0c-ed1d33d43b6d/distribution/dist-dga-32be073c-338b-40df-9842-7cd91a25d960/details?q=} & \scriptsize\url{https://data.gov.au/dataset/ds-dga-19432f89-dc3a-4ef3-b943-5326ef1dbecc/details} & \cite{redlichquantitative} \\ \hline
 
 Earth Engine Data Catalogue & The public data catalogue of Earth Engine encompasses a range of standard raster datasets related to Earth science. \cite{clarke2016multiscale} utilised a public GNSS track from a hike on the Tennessee Valley trail in the Golden Gate National Recreation Area, California, retrieved from the user-contributed data for Google Earth. The track consists of 360 points specified in geographic decimal degrees on the WGS84 coordinate system. & \scriptsize\url{https://developers.google.com/earth-engine/datasets/catalog} & \cite{clarke2016multiscale} \\ \hline
 Deaths recorded in Lawrence, Massachusetts from 1911 to 1913 & This repository, provided by \cite{broen2021measuring}, contains geocoded data on all deaths that occurred in Lawrence, MA, from 1911 to 1913. Additionally, it provides the code for applying various spatial perturbations and generating descriptive spatial statistics. & \scriptsize\url{https://github.com/broenk/Spatial_Perturbation} & \cite{broen2021measuring} \\ \hline
 GIS data for the city of Lawrence, Massachusetts & This repository provides GIS data for the city of Lawrence, Massachusetts. & \scriptsize\url{https://data.mass.gov/browse/gis} & \cite{broen2021measuring} \\ \hline
 GIS data for Orange County, North Carolina & This repository provides GIS data for Orange County, North Carolina. 
\cite{allshouse2010geomasking} employed the E911 database for Orange County, North Carolina (Orange County GIS Division 2007), which includes census block groups exhibiting diverse population densities.& \scriptsize\url{https://www.orangecountync.gov/1610/Land-Records-GIS-Services} & \cite{allshouse2010geomasking} \\ \hline
 Bike sharing dataset (NY, USA) & This repository provides citi bike trip data, including details on their riding locations, preferred timing, ride distances, popular stations, and the days of the week with the highest ride frequency. & \scriptsize\url{https://citibikenyc.com/system-data} & \cite{hasan2017effective} \\ \hline
  Austrian statistical data & The platform data.statistik.gv.at provides various datasets sourced from the official Austrian statistical data dictionary. These datasets adhere to the Principles Of Open Data, ensuring their accessibility in machine-readable formats. In the study by \cite{seidl2015spatial}, energy consumption data for each household in the Hermagor district of Carinthia in southern Austria was utilised. This dataset encompasses 1945 residential records, represented by the centroids of georeferenced buildings supplied by the respective communities in the district. & \scriptsize\url{https://www.statistik.at/datenbanken/statcube-statistische-datenbank} & \cite{seidl2015spatial} \\ \hline
  Open platform for research data (OSF) & OSF serves as a freely accessible and open platform designed to facilitate research and foster collaboration by providing access to projects, data, and materials that could be beneficial to the research process.
  \begin{itemize}[left=0pt]
     \item[-] Synthetic clusters, addresses, and masked data provided by \cite{swanlund2020street}: \url{https://osf.io/6uqkv}: 
     \item[-] Simulated data provided by~\cite{ wang2022exploratory}: \url{https://osf.io/hxd5w/} 
 \end{itemize}& \scriptsize\url{https://osf.io/search?resourceType=File}& \cite{swanlund2020street, wang2022exploratory} \\ \hline
  Figshare & Figshare is an open cloud-based platform used for managing, sharing, and discovering research outputs, including data, figures, code, presentations, and other types of scholarly content.
  \begin{itemize}[left=0pt]
     \item[-] A synthetic address-level dataset provided by~\cite{lin2023geo}: \url{https://figshare.com/articles/dataset/Geo-indistinguishable_masking_Enhancing_privacy_protection_in_spatial_point_mapping/23632443} 
 \end{itemize}& 
 \scriptsize\url{https://osf.io/search?resourceType=File}& 
 \cite{lin2023geo} 
 \\ \hline
\end{longtable}
\end{footnotesize}

\subsubsection{Implementations}
~\\
\noindent This dimension specifies the applications and prototypes that have been developed and implemented for various geoprivacy techniques in order to conduct experimental evaluation.
Some solutions proposed in the literature provide only algorithmic details, but they have not been evaluated experimentally, or no details about their implementation have been published. Table \ref{tab:geomasking_tools} 
provides a brief description of existing tools implementing different
geomasking techniques. 

\begin{table*}[!t]
\caption{Different geomasking tools that have been implemented to ensure geoprivacy.}
\label{tab:geomasking_tools}
\begin{footnotesize}
\begin{tabular}{|p{2.2cm}|p{8.5cm}|>{\raggedright\arraybackslash}p{2cm}|}
\hline
\textbf{Geoprivacy Tool} & \textbf{Description} & \textbf{Repository link} \\ \hline
\textbf{GeoPriv}~\cite{geoprivplugin} &
  A QGIS 3+\footnote{\url{https://qgis.org/en/site/}} plugin that offers 
  three location privacy methods (i.e., Spatial Clustering \cite{zurbaran2017voka}, 
  NRandK~\cite{zurbaran2018nrand}, and Laplace Noise (or Geo indistinguishability) \cite{andres2013geo}). These methods are mainly designed for location-based 
  services and focus on Vector Layers with Point geometries, utilising 
  obfuscation and k-anonymity. & 
  \scriptsize\url{https://github.com/Diuke/GeoPriv} \\ \hline
  
\textbf{MaskMy.XYZ}~\cite{swanlund2020maskmy} &
  A web application designed for easy geographic masking. 
  This tool uses donut geomasking and includes metrics for privacy protection 
  and information loss, enabling users to adjust masking parameters 
  iteratively based on these metrics. & 
  \scriptsize\url{https://maskmy.xyz/} \\ \hline
  
\textbf{PrivyTo}~\cite{mckenzie2022privyto} &
  This tool is designed to protect privacy when sharing personal location 
  data. Users can generate multiple obfuscated views using various geomasking
  techniques (i.e., random perturbations), tailoring each for specific recipients. 
  The application utilises encryption for secure data transfer. & 
  \scriptsize\url{https://privyto.me/site} \\ \hline
  
\textbf{Privy}~\cite{ajayakumar2019addressing} &
  A geomasking tool mainly designed for healthcare. It transfers
  key location information to a different geographic location, analysing
  the data, and finally returns the results to the original location. 
  Privy relocates coordinates using a combination of random translations 
  and rotations while maintaining the connectivity 
  of spatial point data. & 
  \scriptsize\url{https://github.com/ghhlab/confidentiality} \\ \hline
  
\textbf{MapSafe}~\cite{sharma2023mapsafe} &
  A web-based application to share encrypted geospatial information
  that uses donut masking and hexagonal binning for dataset obfuscation, 
  followed by a multi-level encryption scheme. This tool enables data owners 
  to share data at their preferred level of detail. &
  \scriptsize\url{https://www.mapsafe.xyz/} \\ \hline
  
\textbf{StppSim}~\cite{adepeju2023stppsim} &
  A tool designed for synthesising fine-grained spatiotemporal 
  crime records. This tool allows to generate synthetic data from 
  a sample dataset or based on user-provided spatiotemporal 
  characteristics. &
\scriptsize\url{https://github.com/cran/stppSim}
\\ \hline
  
\textbf{GeoMasker}~\cite{chen2017balancing} &
  A geospatial tool designed to 
  balance the protection of geoprivacy and the accuracy of spatial 
  patterns within a GIS. The tool employs carefully calibrated 
  geomasking parameters, including grid size (GS), estimated K-anonymity (K), and D statistics, to mask the residential 
  locations within GIS software. GeoMasker is valuable for health analytics, enabling secure information processing and sharing while preserving 
  privacy and spatial patterns. & 
  \scriptsize\url{http://idv.sinica.edu.tw/tachien/geomasker} \\ \hline
\end{tabular}%
\end{footnotesize}
\end{table*}
\subsubsection{Application Domains}
~\\
Geospatial information, spanning a diverse range of location-based data, plays a pivotal role across various application domains, including but not limited to disaster management, health analytics, urban planning and management, and agriculture. These distinct sectors are explored in greater detail in the subsequent subsections.


\paragraph{Disaster Management}

~\\
Geospatial technologies play a pivotal role in disaster management, offering a
multifaceted toolkit that aids in various aspects of the disaster lifecycle. 
%
Geospatial technologies significantly enhance decision-making, resource allocation,
and coordination across all stages of disaster management~\cite{ghosh2023gis}. 
Early warning systems are facilitated through the integration of real-time data,
such as weather patterns and seismic activities, while crowdsourced data from
mobile applications and social media supplement geospatial information, providing
valuable real-time insights during disasters. The interoperability of geospatial tools supports collaboration among various agencies, and simulation tools aid in training and capacity building for disaster management personnel. 
Additionally, the integration of geospatial technologies with IoT devices 
enhances real-time monitoring of critical infrastructure, contributing to 
the overall preparedness and resilience of communities in the face of 
disasters~\cite{9032921}.

The utilisation of geospatial technologies, 
such as remote sensing and satellite imagery, poses a risk of directly or 
indirectly observing private property and capturing sensitive personal 
information without the possibility of informed consent. In situations where
visual data is collected and disseminated, individuals may be exposed to privacy
violations, placing them in vulnerable positions. Moreover, the absence of 
robust privacy and data protection standards, especially in the context of data sharing, heightens the potential misuse of geospatial data. This ethical
dilemma underscores the importance of addressing privacy issues to ensure 
responsible and transparent practices in the use of GIS for emergency 
management \cite{kicior2023,9533265}.

\paragraph{Health Analytics}
~\\
Geospatial technology plays a crucial role in healthcare by providing spatially explicit
information about environmental exposures, lifestyle patterns, and disease risks. 
Over the last three decades, advancements in Earth observations, spatial data quality,
and software usability have enabled the integration of geospatial tools in various
health studies~\cite{faruque2022geospatial,bhoda_geospatial_health,shah2023revolutionizing}. 
This technology not only contributes to a comprehensive understanding
of the complex interplay between genes, lifestyle, and environment but also extends
its impact on crucial aspects of public health. Geospatial data serves as a powerful
tool for disease surveillance, allowing the collection and analysis of data to 
monitor disease outbreaks and inform timely responses~\cite{faruque2022geospatial}.
It aids in identifying 
at-risk populations, guiding the allocation of resources during 
outbreaks~\cite{shah2023revolutionizing}. 

Furthermore, geospatial technology enhances health promotion efforts by offering
detailed information on the accessibility of healthcare services, helping 
policymakers make informed decisions about facility placement to ensure equitable
access for all~\cite{sahana2022geospatial}. In addition, it plays a key role in 
evaluating and improving healthcare delivery by providing insights into the quality
of care across different facilities. While the integration of geospatial technology
into healthcare raises privacy concerns that need addressing, its potential to 
optimise healthcare systems globally and contribute to improved public health 
outcomes is undeniable~\cite{faruque2022geospatial}.
It is crucial for the healthcare industry to navigate the challenges and ethical
considerations associated with geospatial intelligence, ensuring the responsible
use of location-based data while harnessing its power to revolutionise healthcare
delivery and management. 
\paragraph{Urban Planning and Management}
\medskip
~\\
Geospatial technology plays a pivotal role in shaping smart cities, providing
a robust foundation for enhanced urban management. Through the integration of
diverse geospatial tools such as GIS, GPS, remote sensing, and 3D modeling, 
smart cities can seamlessly collect real-time data on various facets of 
their environment, infrastructure, and citizen activities.
Geospatial technology facilitates a spectrum
of applications, from urban planning and transportation management to 
emergency services and environmental monitoring. For instance, it enables
the creation of dynamic maps that aid in economic development, event management,
and route planning. Moreover, this technology supports initiatives like 
renewable energy mapping, hazard identification, and infrastructure planning,
contributing to the sustainable growth and resilience of modern urban 
landscapes \cite{padode2015role}.


Balancing the benefits of smart city technologies with the imperative 
to protect individual privacy becomes crucial. This necessitates thoughtful
considerations regarding data governance, transparency, user consent, 
and ethical use of surveillance technologies to prevent unwarranted 
intrusions into individuals' private lives. Efforts to address these 
privacy challenges must involve setting balanced default parameters, 
providing users with control over data settings, and promoting aggregate
data processing to ensure anonymity, all while maintaining the delicate 
equilibrium between technological innovation and individual privacy 
rights~\cite{fabregue2023privacy}.
\paragraph{Agriculture}
\medskip
~\\
Geospatial technology plays a pivotal role in 
agriculture. 
Through the utilisation of remotely sensed images acquired from
satellites, aerial platforms, and GPS-tagged drones, geospatial technology
facilitates the mapping and monitoring of various agricultural parameters,
including soil moisture, nutrient availability, crop types, growth stages,
and disease stress. 
The integration of geospatial technology into precision agriculture not
only supports sustainable development goals but also aids in conserving 
biodiversity, mitigating environmental impacts, and promoting sustainable
land management practices 
worldwide \cite{pandey2023highlighting,obi2023applications}.

Privacy concerns in agriculture arise from the widespread adoption of 
geospatial technology and digital farming practices. The use of GPS-linked
data in digital farming, aimed at achieving sustainable agriculture, poses
risks of re-identification and disclosure of private information due to the
spatial nature of agricultural data. To mitigate these concerns, spatial 
anonymisation, obfuscation, and geomasking methods are 
used~\cite{nowbakht2022geoprivacy}. 

\section{Existing literature on geomasking techniques}
\label{sec:survey}

In this section, a comprehensive survey of recent geomasking techniques is 
conducted, examining the methods employed to preserve geospatial privacy 
while maintaining data utility and accuracy. We provide the details of 
the existing literature following the categorisation in 
Section~\ref{sec:geomasking-techniques}.

\subsection{Random perturbation}

\begin{table*}[t!]
\centering
\caption{Existing literature with masking using random perturbation techniques.}
\label{tab:random-pert-lit}
\begin{footnotesize}
\begin{tabular}{|l|l|l|l|} 
\hline
\multicolumn{1}{|l|}{\multirow{5}{*}{\begin{tabular}[c]{@{}l@{}}Density adaptive \\ displacement\end{tabular}}} &
  \multirow{3}{*}{\begin{tabular}[c]{@{}c@{}}Uniform population \\ density distribution\end{tabular}} &
  \begin{tabular}[c]{@{}c@{}}Circular masking with fixed radius\end{tabular} & \cite{kwan2004protection}
   \\ \cline{3-4} 
\multicolumn{1}{|l|}{} &
   &
  \begin{tabular}[c]{@{}c@{}}Circular masking with random radius\end{tabular} &
   \cite{armstrong1999geographically,kwan2004protection}\\ \cline{3-4} 
\multicolumn{1}{|l|}{} &
   &
  \begin{tabular}[c]{@{}c@{}}Random directions within annual (Donut masking)\end{tabular} &
    \cite{stinchcomb2004procedures, allshouse2010geomasking, hampton2010mapping}\\ \cline{2-4} 
\multicolumn{1}{|l|}{} &
  \multirow{2}{*}{\begin{tabular}[c]{@{}c@{}}Heterogeneous population \\ density distribution\end{tabular}} &
  \begin{tabular}[c]{@{}c@{}}Spatially adaptive random perturbation (SARP)\end{tabular} &
   \cite{lu2012considering} \\ \cline{3-4} 
\multicolumn{1}{|l|}{} &
   &
  \begin{tabular}[c]{@{}c@{}}Weighted random perturbation (WRP)\end{tabular} &
   \cite{allshouse2010geomasking}\\ \hline
\multicolumn{2}{|l|}{\multirow{3}{*}{\begin{tabular}[c]{@{}l@{}}Gaussian deviation for \\ displacement magnitude\end{tabular}}} &
  Gaussian displacement &
   \cite{cassa2008re,fanshawe2011spatial,zandbergen2014ensuring}\\ \cline{3-4} 
\multicolumn{2}{|l|}{} &
  \begin{tabular}[c]{@{}c@{}}Bimodal Gaussian displacement\end{tabular} &
   \cite{cassa2006context,zandbergen2014ensuring}\\ \cline{3-4} 
\multicolumn{2}{|l|}{} &
  \begin{tabular}[c]{@{}c@{}}Density based Gaussian spatial skew\end{tabular} &
   \cite{cassa2006context}\\ \hline
\multicolumn{2}{|l|}{\begin{tabular}[c]{@{}l@{}}Multiple risk factor adaptive displacement\end{tabular}} &
  Triangular Displacement & \cite{murad2014protecting}
   \\ \hline
\end{tabular}%
\end{footnotesize}
\end{table*}

Table~\ref{tab:random-pert-lit} outlines different papers that proposed or 
utilised a random perturbation technique. Kwan et al.~\cite{kwan2004protection}
investigate the use of circular masking with a fixed and random radius $r$. 
According to the investigation, a larger $r$ for the same mask leads to more 
deviation from the pattern obtained from the unmasked data. Further, the study 
suggests that circular and random radius masks, in general, yield better 
results than the weighted masks with the same $r$~\cite{allshouse2010geomasking}.

Stinchcomb~\cite{stinchcomb2004procedures} was the first to explore 
donut masking for displacing points in random directions within an 
annulus. However, in cases where population density is not evenly 
distributed, donut masking could lead to revealing information about
individuals. To mitigate this, Allshouse et al.~\cite{allshouse2010geomasking}
suggest tripling the minimum displacement area within the donut 
for heterogeneous populations, ensuring privacy preservation with 
less than a 1\% error rate.

Further, Hampton et al.~\cite{hampton2010mapping} examine the donut 
method for disease mapping and compares its performance with other 
geomasking techniques (i.e., aggregation and random perturbation) in 
terms of cluster detection and patient geoprivacy protection. Their 
investigation shows that the donut method and random perturbation 
outperformed aggregation in measuring hot spot/clustered areas. 
However, the donut method provides a higher level of privacy 
preservation compared to the random perturbation, with only a slight reduction in cluster detection performance, especially 
in areas where individual privacy is at a higher risk.

Lu et al.~\cite{lu2012considering} explore a masking technique based
on Spatially adaptive random perturbation (SARP). 
As we have discussed in 
Section~\ref{sec:geomasking-techniques}, the SARP technique defines the 
perturbation zone based on the actual distribution of residential 
addresses (risk location) instead of people (risk population). 
It employs "donut-shaped" spatially varied perturbation zones rather 
than "pancake-shaped" which allows a point to be moved to a nearby 
location within its immediate surroundings. The study findings show
that SARP geomasking, utilising actual street addresses, offers better
privacy protection compared to geomasking based on population size. 
While the SARP techniques do not significantly alter clustering 
patterns on a global scale, the geomasked data tends to exhibit 
greater clustering compared to the real distribution.

Cassa et al.~\cite{cassa2006context}, Zandbergen~\cite{zandbergen2014ensuring}, 
and Cassa et al.\cite{cassa2008re} investigate the masking methods based on 
Gaussian displacement. However, these studies highlight that the main 
challenge of Gaussian displacement lies in the limited point displacement due
to its concentration around the mean. As highlighted by\cite{zandbergen2014ensuring},
donut masking and bimodal Gaussian displacement are very similar in terms of 
the general area where masked locations are placed relative to the original locations.

Murad et al.~\cite{murad2014protecting} investigate the use of triangular 
displacement as a masking technique for geodata. The authors suggest a method that 
evaluates various risk factors to determine the appropriate displacement distances 
for individual location points. 
The method considers the sensitivity level of the data, adding specific 
displacement distances for low and high sensitivity information. 
Quasi-identifiers, which pose a
combined risk when aggregated, prompt the algorithm to add additional 
distance to the displacement distance, while grouping these quasi-identifiers 
until a preset number of individuals are in each group. 
The algorithm then
considers the intended release of the dataset, 
differentiating between public and restricted access. 
The resulting 
displacement distance, treated as a range with lower and upper limits
($R_{Lower}$, $R_{Upper}$), is employed in the Triangular Displacement 
method. This method, based on the Pythagorean Equation, 
introduces a random yet controlled shift in the original location points, 
providing a versatile and context-aware approach to spatial data masking. 

\subsection{Masking with pre-defined locations}

Zhang et al.~\cite{zhang2017location} was the first to use location swapping as a masking technique for geospatial data in the health analytic domain. This technique offers a more realistic scenario for displacement location selection. However, as the authors highlighted, if an attacker is able to determine the buffer distance used to relocate a point, it might be possible to re-identify the original locations. The same authors also explored the location swapping with donut masking to set a minimum distance preventing points from being swapped with any locations within this specified range. The authors employed an internal buffer size half that of the external buffer.

Swanlund et al.~\cite{swanlund2020street} introduce street masking, which uses 
a road network as a basis for masking. The authors use OpenStreetMap road network
to replace the original location with coordinates sampled from a set within a 
predefined radius. As the study suggests, street masking performs similarly to 
population-based donut geomasking with regard to privacy protection, 
achieving comparable k-anonymity values at similar median displacement distances. 

Geo-indistinguishability, as proposed by Andrés et al.~\cite{andres2013geo}, is a privacy notion proposed for preserving privacy in location-based services. It offers stronger protection than spatial k-anonymity by safeguarding against adversaries with any current or future side information. The notion is formalised as a user enjoying $\epsilon r$-privacy within a radius $r$, where $\epsilon$ determines the level of privacy per unit distance, with smaller values of $\epsilon$ providing higher privacy protection. It achieves privacy protection by relocating individual point locations, ensuring that within a certain radius, any two points yield similar locations after relocation.

Lin~\cite{lin2023geo} introduces geo-indistinguishable masking as a robust solution to safeguard privacy in spatial point mapping.  By leveraging the concept of geo-indistinguishability, the approach ensures strong privacy protection while preserving spatial point patterns through a two-step process. Initially, masking areas are generated around each spatial point, comprising a set of candidate locations for potential relocation. Subsequently, an optimisation model is employed to strategically select the optimal candidate location within each masking area, minimising displacement distance while adhering to the stringent privacy criterion of geo-indistinguishability. While sharing similarities with location swapping methods, which also relocate spatial points within masking areas, geo-indistinguishable masking distinguishes itself by prioritising optimal candidate selection over random assignment, thereby offering superior privacy guarantees. Despite its computational complexity and potential limitations regarding temporally correlated spatial points and risk of false identification, the method represents a significant advancement in the realm of privacy-preserving spatial analysis, essential for upholding individual privacy rights in an era of burgeoning geospatial data utilisation.

\subsection{Affine transformation} 

Recently, Wang et al.~\cite{wang2022exploratory} conduct a study that 
investigates the use of translation, rotating, and scaling techniques for 
geomasking. The authors highlight that the coordinates of the original records
can be easily recovered from the geomasked dataset if the attacker knows the 
transformation matrix, and even the identification of only a few points could make it 
possible to re-identify the entire point pattern.

Leitner and Curtis were the first to explore flipping as a geomasking 
technique~\cite{leitner2004cartographic}. The authors state that most 
variants (e.g. flipping horizontally and flipping vertically) of the flipping method 
are designed for publishing maps and not for releasing the coordinates~\cite{leitner2006first}.
Broen et al.~\cite{broen2021measuring} investigate the impact of horizontal shear
as a masking technique. According to the authors, horizontal shear confers the 
greatest anonymity using the k-anonymity metric and maintains some spatial patterns
compared to other affine transformation techniques. However, the method is not secure, 
as points can be trivially re-identified if the angle of the shearing can be determined.

\subsection{Spatial aggregation and Clustering}
 Table~\ref{tab:aggre-clust-lit} outlines the literature that
uses spatial aggregation and clustering techniques for masking geodata. 


\begin{table*}[t!]
\centering
\caption{Existing literature with Spatial aggregation and clustering}
\label{tab:aggre-clust-lit}
\begin{footnotesize}
\begin{tabular}{|lll|l|}
\hline
\multicolumn{1}{|l|}{\multirow{4}{*}{Grid based aggregation}} &
  \multicolumn{1}{l|}{\multirow{2}{*}{Fixed size grid cells}} &
  Grid line masking & \cite{leitner2004cartographic,gupta2020preserving}
   \\ \cline{3-4} 
\multicolumn{1}{|l|}{} & \multicolumn{1}{l|}{} & Grid center masking                                                         & \cite{leitner2004cartographic,gupta2020preserving} \\ \cline{2-4} 
\multicolumn{1}{|l|}{} &
  \multicolumn{1}{l|}{Varying size grid cells} &
  \begin{tabular}[c]{@{}l@{}}Quadtree hierarchical \\ geographic data structure\end{tabular} &
   \cite{lagonigro2017quadtree}\\ \cline{2-4} 
\multicolumn{1}{|l|}{} & \multicolumn{2}{l|}{Military Grid Reference System (MGRS)}                                          &  \cite{clarke2016multiscale}\\ \hline
\multicolumn{2}{|l|}{\multirow{6}{*}{\begin{tabular}[c]{@{}l@{}}Spatial Tessellation based techniques\end{tabular}}} &
  Voronoi masking & \cite{seidl2015spatial}
   \\ \cline{3-4} 
\multicolumn{2}{|l|}{}  & \begin{tabular}[c]{@{}l@{}}Adaptive areal elimination (AAE)\end{tabular} &  \cite{kounadi2016adaptive}\\ \cline{3-4} 
\multicolumn{2}{|l|}{} & \begin{tabular}[c]{@{}l@{}}Adaptive areal masking (AAM)\end{tabular}     &  \cite{charleux2020true}\\ \cline{3-4} 
\multicolumn{2}{|l|}{}                         & \begin{tabular}[c]{@{}l@{}}Adaptive voronoi  masking (AVM)\end{tabular}   &  \cite{polzin2020adaptive}\\ \cline{3-4} 
\multicolumn{2}{|l|}{}                         & \begin{tabular}[c]{@{}l@{}}RDV masking\end{tabular}   &  \cite{gupta2020preserving}\\ \cline{3-4} 
\multicolumn{2}{|l|}{}  & Zip4-codes aggregation & \cite{ajayakumar2023utility} \\ \hline
\multicolumn{3}{|l|}{Differentially private clustering}                                                                      &  \cite{lu2020differentially,su2017differentially,balcan2017differentially,ghazi2020differentially,cohen2021differentially,zhang2021mpdp,jones2021differentially,ni2021utility,chen2023improved}\\ \hline
\multicolumn{3}{|l|}{Aggregation and perturbation}                                                                           &  \cite{houfaf2021geographically}\\ \hline
\end{tabular}%
\end{footnotesize}
\end{table*}

As explored by~\cite{leitner2004cartographic,gupta2020preserving}, grid masking disparate
every original location data point to uniform grid cells. However, grid masking
disrupts the original data points' pattern and reduces the data utility for further 
research and analysis. Lagonigro et al.~\cite{lagonigro2017quadtree} propose 
a methodology based on quadtree hierarchical geographic data structures to
create a varying-size grid adapted to local area densities. As a hierarchical 
grid system, it can easily be aggregated into a fixed-size grid on a lower
resolution level. The methodology has been implemented using the R software 
and applied to the Catalonian 2014 population register data. 
Clarke~\cite{clarke2016multiscale} investigates a multiscale geomasking technique, whereby locations are converted to Military Grid Reference System (MGRS) coordinates, providing a unique level of control over the adjusted locations. Despite the computing time required for processing, the method offers new 
possibilities for protecting sensitive geospatial data, overcoming disadvantages
of geometric point masking methods. 

Voronoi masking (VM) is a spatial tessellation technique that can retain the spatial 
integrity of the original data. Seidl et al.~\cite{seidl2015spatial} was the first
to introduce the idea of VM. One advantage of this method is that 
points in neighbouring polygons are displaced to the same position, enhancing
their K-anonymity. However, VM can dislocate some data points to large distances if the area contains scatted positions, which will change spatial patterns. This mask was automated in 
Python for ArcGIS Pro and ArcMap. 

Kounadi and Leitner~\cite{kounadi2016adaptive} propose Adaptive areal elimination (AAE)
which creates areas of a minimum K-anonymity, and then original points are either 
randomly perturbed within the areas or aggregated to the median centers of the areas. 
According to the authors, AAE introduces less spatial error than the donut mask for 
all the measures of spatial error. Polzin~\cite{polzin2020adaptive} integrates the 
principles of AAE and VM to introduce a masking technique called Adaptive voronoi 
masking (AVM). AVM addresses topographical errors by relocating data points to the 
nearest street intersection, minimising false re-identification and illogical 
placements in water bodies or forests.

Gupta and Rao~\cite{gupta2020preserving} propose a method very similar to 
VM called RDV masking. Depending on the region size, for large regions, points 
are moved to the nearest segment of the edges of the polygon. Charleux and Schofield~\cite{charleux2020true} introduce a variation of AAE called 
Adaptive Areal Masking (AAM). This method tries to merge polygons not by the 
largest shared border but with the polygon with the closest centroid. However,
in configurations where the number of points to anonymise is high enough and 
the number of population polygons is low enough, AAE can outperform AAM.
Similar to AAE, anonymisation polygons generated by the AAM method should 
not be disclosed to end users to minimise any potential reidentification attacks. 

Ajayakumar et al.~\cite{ajayakumar2023utility} propose to use Zip4 
codes as a spatial aggregation unit for sharing fine-scale health data. 
Analysing 1.6 million Ohio Zip4 codes, the research reveals Zip4's high 
spatial precision, particularly in urban locales, making it well-suited 
for detailed spatial analysis than common aggregation units. However, the study 
underscores a potential privacy risk, noting that Zip4 centroids have a
higher likelihood of compromising spatial anonymity, with 73\% of 
addresses exhibiting a low spatial k-anonymity value. 

Differential private (DP) clustering focuses on the critical challenge of clustering
sensitive data while preserving individual privacy, which holds significant 
importance in machine learning and data analysis applications. 
Su et al.~\cite{su2017differentially} delve into the limitations of current 
state-of-the-art approaches for k-means clustering, particularly focusing on 
methodologies that employ a single-workload technique. Ghazi et al.~\cite{ghazi2020differentially} 
delve into differentially private clustering, focusing on fundamental clustering 
problems such as Euclidean DensestBall, 1-Cluster, k-means, and k-median.
Cohen et al.~\cite{cohen2021differentially} and Lu and Shen~\cite{lu2020differentially} propose a new approach to 
integrating differential privacy into clustering models, addressing the limitations of existing differentially
private clustering algorithms. Yang et al.~\cite{yang2022k} address the emerging field 
of privacy-preserving data analysis, specifically focusing on constructing a 
privacy-preserving K-means clustering protocol that ensures local privacy for 
user data.

Houfaf-Khoufaf et al.~\cite{houfaf2021geographically} study the combination of
aggregation techniques with perturbation mechanisms. The study also introduces the 
simulated cluster crowding (SCC) method, a novel approach inspired by the 
simulated crowding of bike GPS tracks. SCC generates new random points within
existing clusters' extents to achieve both high k-anonymity and cluster 
preservation. The clusters are initially generated using the DBSCAN algorithm, 
distinguishing between outlier points and those within clusters. Outlier 
points are masked using the bimodal Gaussian perturbation method, while 
points within clusters are masked by generating the same number of new 
points randomly inside the polygon extent of each cluster. Three versions 
of SCC are proposed: SCCv1 generates points uniformly within the cluster
area, SCCv2 introduces buffer areas around initial points to ensure a 
minimum distance, and SCCv3 dilates roads and intersects them with 
clusters to generate fake points. The methods are evaluated based on 
k-anonymity, cluster preservation, and privacy guarantee, demonstrating
their effectiveness in maintaining privacy while preserving the spatial
distribution of clusters.

\subsection{Synthetic data generation}
 Quick and Waller~\cite{quick2018using} 
propose the use of spatiotemporal data analysis methods for generating synthetic data 
to address disclosure risks associated with releasing public-use data in disease 
mapping, where small counts present challenges. The study employs a Bayesian model where
the synthetic data are generated from the resulting posterior predictive distribution, 
preserving different dependencies. Drechsler and Hu~\cite{drechsler2021synthesizing} investigate the 
viability of using synthetic data generation to provide detailed geocoding information
from a large administrative database while maintaining confidentiality. The authors show
that a synthesis strategy based on categorical CART models performs well in addressing the
risk-utility trade-off.

Quick~\cite{quick2021generating} focuses on the dissemination of synthetic data as an 
effective means of making sensitive information publicly available while minimising 
the risk of disclosure. It extends a differential privacy approach to Poisson-distributed 
count data, accommodating population size heterogeneity and incorporating prior information
on event rates. Hong et al.~\cite{hong2022collecting} propose a privacy-preserving 
synthetic data generation framework for addressing privacy concerns associated with 
sharing complete datasets. The method involves synthesising a new privacy-preserving 
dataset through a generative model that captures the probability density function 
of the original dataset's features. The synthetic dataset is designed to be statistically
distinct from the original while preserving data utility.  

Koebe et al.~\cite{koebe2023releasing} propose an innovative microdata dissemination 
strategy for household survey data to balance privacy and data utility. Instead of 
traditional approaches involving perturbed cluster locations, they suggest publishing 
two datasets: the first contains original survey data with limited geographic 
identifiers, suitable for representative-level analysis; the second involves 
synthetically generated survey data with accurate cluster locations, enhancing
proximity-related information for spatial analysis. The proposed strategy significantly
reduces re-identification risks compared to common spatial perturbation methods, 
while maintaining data utility for both non-spatial and spatial analyses. The authors
employ copulas for synthetic data generation, emphasising their computational efficiency, interpretability, and practicality for data producers. The methodology is seen as 
beneficial for expanding the use cases of survey data products, supporting mapping 
initiatives, and ensuring data access while complying with privacy regulations.

\section{Open issues and future directions}
\label{sec:future-directions}

As we have reviewed in Section~\ref{sec:survey}, no geomasking technique is complete and capable of providing high accuracy while ensuring strong privacy against different attacks. Each masking technique has distinct advantages and disadvantages, requiring data custodians to understand these techniques and their capabilities before applying them in practical applications. While most masking literature provides a detailed 
explanation of how to use the masking procedure in general, most masking 
methods lack a corresponding guide for the optimal choice of parameters, 
as these parameter settings need to be changed according to the data that needs
protection. Further, some papers do not take into account different scenarios 
that could occur when their proposed masking techniques are applied in real applications.
For example, a technique such as location swapping could lead to issues when 
no residential addresses are found within the radius that is defined based on the
population density of a region~\cite{fabregue2023privacy}. 

The main question in geomasking research is how to balance the need for
anonymisation with the usefulness of geodata while also considering the 
probability and potential harms of re-identification~\cite{helderop2023unmasking}. In general, geomasking methods can preserve data accuracy but are presumed to impose a high risk of re-identification. As discussed in Section~\ref{sec:geomasking-techniques}, complete anonymisation is impossible if the data is to remain useful. Therefore, research efforts should focus not only on better preventing identification but also on considering the probability of re-identification events and the potential harms of such incidents.

False identification is an emerging concern of applying geomasking methods,
which indicate the linking of the masked data points to incorrect persons 
or households~\cite{seidl2018privacy}. The false identification transferred
the potential negative effects of being identified from the true persons or
households to individuals who were not part of the research. Such 
limitations influence both the disclosure risk and a successful 
investigation of spatial patterns. There are newly developed geomasking 
methods that target addressing this issue, such as adaptive Voronoi 
masking~\cite{kounadi2016adaptive}. Though these techniques 
might be better approaches to visualising a protected version of the 
distribution of a point pattern, however, they are less accurate in detecting
local patterns in location data. Further, these techniques might not be 
applicable to different geodata types that are complex, such as spatiotemporal
stamps of a user in a social network.

As we have discussed in Section~\ref{sec:geomasking-techniques}, many geomasking
techniques could not provide provable privacy guarantees as their underlying 
methodologies only contain non-provable privacy techniques, such as 
k-anonymity~\cite{wang2022exploratory, broen2021measuring}. Techniques such as differential privacy have been used in location-based services~\cite{yan2022} 
for providing strong and provable privacy guarantees when answering queries, but have limited adoption and utilisation in the masking of geodata~\cite{harris2020}. For example, to improve geodata anonymisation and align with the relevant risk factors, it may have been more appropriate to 
add random noise to the geographic information of routes mapped by just a few 
individuals, or exclude routes mapped by very few individuals (e.g., publish
the number of routes in the area, but not the exact paths). 

In the geomasking literature, each paper tends to use different datasets and
different measures to evaluate their proposed methodologies. As shown by 
Redlich~\cite{redlichquantitative}, the choice of different accuracy and 
privacy metrics influence the evaluation of the geomasking methods, leading 
to different conclusions. The utilisation of different experimental settings in geomasking literature could potentially make the comparison of different geomasking techniques difficult. Thus, having a benchmark setting, including a set of datasets, accuracy measures, and privacy measures, would allow for more accurate quantification of the performance of different geomasking methods.

\begin{table}[!t]
\centering
\caption{Tools designed to integrate large language models with 
geospatial data.}
\label{tab:llm-geospatial}
\begin{footnotesize}
\begin{tabular}{|p{2cm}|p{11cm}|}

\hline
\textbf{Tool} & \textbf{Description} \\ \hline
\textbf{IBM Geospatial foundation model} ~\cite{ibm_geospatial_model}   & In collaboration with NASA, IBM has developed the first-ever foundation 
model for analysing geospatial data, watsonx.ai model, to convert satellite
data into high-resolution maps. The model aims to assist in estimating 
climate-related risks, monitoring forests for carbon-offset programs, 
and creating predictive models for climate change adaptation. This model, based on NASA’s Harmonised Landsat Sentinel-2 data, is trained
to understand satellite images, recognising features like historic 
floods and fire burn scars.
  \\ \hline
\textbf{ChatGeoPT}~\cite{strong_chatgeopt_2023} &
ChatGeoPT, a Geospatial AI Assistant, is a prototype that explores the integration of LLMs into geospatial 
data science workflows. The application accepts natural language 
prompts and translates them into queries for the OpenStreetMap (OSM).
ChatGeoPT is seen as a preliminary step toward the development of a 
more extensive Large Earth Observation Model (LEOM), aiming to enable 
intelligent interactions and summarisations of geospatial data from 
diverse sources, including satellite imagery and weather data, 
extending beyond the capabilities of OpenStreetMap.
\\ \hline

\textbf{MapsGPT}~\cite{mapsgpt} &
MapsGPT simplifies the process of creating interactive maps, eliminating
the need for extensive coding skills and making it accessible to everyone. 
Unlike traditional GIS, MapsGPT allows users to generate customised maps 
effortlessly by inputting a simple sentence. MapsGPT provides a versatile
solution for discovering and curating locations, making it easy to plan 
activities, find essential stores, and explore a new city.
\\ \hline

\textbf{LLM-Geo}~\cite{li2023autonomous} &
LLM-Geo represents an innovative paradigm known as Autonomous GIS, 
introducing an AI-powered geographic information system that leverages LLMs 
like ChatGPT to transform spatial analysis. LLM-Geo seeks to automate 
spatial data collection, analysis, and visualisation by harnessing the 
natural language understanding, reasoning, and coding capabilities of LLMs. 
\\ \hline

\textbf{GEoLM}~\cite{li2023geolm} &
GEoLM, a geospatially grounded language model, enhances natural language 
understanding by integrating valuable geospatial information from extensive
databases like OSM. GEoLM connects linguistic details in 
textual corpora with geospatial data retrieved from geographical databases. 
GEOLM stands as a versatile tool for various geographically related language
understanding tasks, with future plans to extend its applications to 
question-answering and auto-regressive tasks.
\\ \hline

\textbf{GFM}~\cite{mendieta2023towards} &
GFM develops geospatial foundation models for applications like agriculture, 
urban planning, and disaster response. The authors introduce GeoPile, a compact
and diverse dataset sourced from various remote sensing platforms, to enhance 
feature diversity during pretraining. They explore the potential of continual pretraining from readily available ImageNet-22k models. 
\\ \hline
\end{tabular}%
\end{footnotesize}
\end{table}

\subsection{Integration of large language models (LLMs) with Geospatial data}

In recent years, substantial progress has been achieved in the development
of foundational models. Notably, the integration of these foundational 
models with geospatial data, referred to as \textit{GeoAI Foundation Models}, 
has demonstrated significant potential across various
applications, ranging from geographic question answering and remote sensing 
image understanding to map generation and location-based 
services~\cite{zhang2023geogpt}. 
Nevertheless, it is crucial to acknowledge that the development and application
of GeoAI foundation models introduce significant privacy and security risks,
an aspect that has not yet received comprehensive attention to date. 

Rao et al.\cite{rao2023building} examine the privacy and security risks 
associated with the entire lifecycle of GeoAI foundation models. The study 
identifies a range of potential privacy and security risks in different 
phases of GeoAI foundation model development. The authors highlight that leveraging such foundational models in geospatial research poses challenges, particularly in balancing privacy and utility, mitigating geographic biases, and addressing cross-modal privacy risks.

Several other studies have explored the use of foundation models in geospatial
data analysis. Mooney et al.~\cite{mooney2023towards} investigate the 
performance of ChatGPT in a real Geographic Information Systems (GIS) exam. 
The study identifies challenges in teaching spatial concepts to LLMs, mainly 
due to the limitations in handling multimodal content.
Tao and Xu~\cite{tao2023mapping} explore the potential of leveraging ChatGPT for the task 
of map design. The authors conduct two sets of experiments, testing the 
generation of thematic maps through specified prompts and the creation of mental
maps purely from textual descriptions of geographic spaces. Though ChatGPT's reliance
on textual input and output, the authors show that ChatGPT holds the potential to
revolutionise map design processes, particularly in educational and research contexts,
while emphasising the need for cautious consideration of its current limitations 
and validation of data sources. 

Similarly, Zhang et al.~\cite{zhang2023geogpt} propose GeoGPT, a framework that
integrates the semantic understanding capabilities of LLMs. GeoGPT utilises the
Langchain framework to connect the LLM with a pool of GIS tools, allowing it to 
choose and execute appropriate tools for tasks like geospatial data collection, 
spatial analysis, and mapping. Zhang et al.~\cite{kang2023ethics} explore the ethical dimensions
of leveraging generative AI in the domain of cartography. Focusing specifically 
on maps generated by LLMs, the researchers establish an open-source dataset 
containing both AI-generated and human-designed maps. The authors show that the maps generated by the LLM lack geographic coverage and diversity. However, they highlight that such AI tools could be valuable in research settings where real geodata is limited. Table~\ref{tab:llm-geospatial} highlights some other tools specifically 
designed to integrate LLMs with geospatial data.

\section{Conclusion}
\label{sec:conclusion}

In this paper, we present a survey of historical and current state-of-the-art masking techniques for geodata. 
We proposed a taxonomy that serves as a comparison and analysis tool for geomasking techniques. 
In this taxonomy we identified three main dimensions: privacy techniques, evaluation measures, and practical aspects. These dimensions enabled us to characterize geomasking techniques and generate a taxonomy of such techniques. 
Through this taxonomy, we identified various shortcomings in current approaches to geomasking, suggesting several future research directions in this field.
An area for future research involves investigating 
how large language models (LLMs) can be employed to analyse and enhance geomasking 
techniques. To date, no studies have delved into the potential integration of 
LLMs for this purpose, making it a promising area for future exploration
and improvement.
Solving these open research questions is a core requirement to make geomasking 
applicable for practical applications.


\bibliographystyle{ACM-Reference-Format}
\bibliography{survey}

\end{document}